\documentclass[a4paper, 11 pt]{article}
\pdfoutput = 1

\usepackage{float}
\usepackage{jheppub, bm}
\usepackage[utf8]{inputenc}
\usepackage{xcolor}

\usepackage{tikz}
\usetikzlibrary{matrix,shapes,fit,tikzmark,calc}

\usepackage{graphicx}
\usepackage{subcaption} 

\usepackage{amsmath}
\usepackage{slashed} 
\usepackage{physics} 
\usepackage[capitalise]{cleveref} 

\usepackage{hyperref}

\usepackage[export]{adjustbox}
\newcommand{\nn}{\nonumber}

\allowdisplaybreaks[2]
	
\title{Nuclear modified transverse momentum dependent parton distribution and fragmentation functions}
	
\author[a]{Mishary Alrashed,}
\author[a,b,c]{Zhong-Bo Kang,}
\author[d]{John Terry,}
\author[e,f]{Hongxi Xing,}
\author[a]{and Congyue Zhang}

\affiliation[a]{Department of Physics and Astronomy, University of California, Los Angeles, California 90095, USA}
\affiliation[b]{Mani L. Bhaumik Institute for Theoretical Physics, University of California, Los Angeles, California 90095, USA}
\affiliation[c]{Center for Frontiers in Nuclear Science, Stony Brook University, Stony Brook, New York 11794, USA}
\affiliation[d]{Theoretical Division, Los Alamos National Laboratory, Los Alamos, New Mexico 87545, USA}
\affiliation[e]{Guangdong Provincial Key Laboratory of Nuclear Science, Institute of Quantum Matter, South China Normal University, Guangzhou 510006, China}
\affiliation[f]{Guangdong-Hong Kong Joint Laboratory of Quantum Matter, Southern Nuclear Science Computing Center, South China Normal University, Guangzhou 510006, China}

\emailAdd{malrashed@physics.ucla.edu}
\emailAdd{zkang@ucla.edu}
\emailAdd{jdterry@lanl.gov}
\emailAdd{hxing@m.scnu.edu.cn}
\emailAdd{maxzhang2002@g.ucla.edu}

\abstract{In this study, we extend our previous global analysis of nuclear-modified transverse momentum distribution functions (nTMDs) to also consider the nuclear-modified collinear fragmentation function. Our methodology incorporates the global set of experimental data from both Drell-Yan production and Semi-Inclusive Deep Inelastic Scattering. Through a comprehensive global extraction of these distributions, we demonstrate the effectiveness of this extension by strongly describing the entire global dataset. A focal point of this paper is the impact of recent Jefferson Lab measurements. Most notably, to simultaneously describe experimental data at Jefferson Lab and HERMES we find that it is necessary to introduce a parameter which accounts for the non-perturbative scale evolution of the nTMDs. Additionally, we assess the kinematic coverage of the experimental data and provide insights into experimental opportunities at Jefferson Lab, future Electron-Ion Colliders, RHIC, and the LHC. These opportunities have the potential to significantly enhance and refine global analyses of nuclear-modified TMDs, contributing to a deeper understanding of the structure of cold nuclear matter.}
\begin{document}
\preprint{LA-UR-23-33438}
\maketitle

\section{Introduction}\label{Introduction}
Unveiling the femtoscale structure of both free and bound nucleons is a key objective in the field of nuclear physics \cite{Accardi:2012qut,AbdulKhalek:2021gbh,Anderle:2021wcy}. To address these challenges, we are forced to explore the longstanding question of how the ultra-violet behavior of the strong force, which is governed by perturbative QCD, correlates with its infrared dynamics. Over the past several decades, the community has been deeply engaged in connecting these disparate regions by extensively studying Transverse Momentum Dependent Parton Distribution Functions and Fragmentation Functions (TMD PDFs and TMD FFs). These correlation functions relate the three-dimensional partonic and hadronic momenta, serving as a vital tool in understanding the intricate behaviors of the strong force across various energy scales and deepening our understanding of the three-dimensional structure of matter and the mechanism of hadronization.

TMD PDFs and TMD FFs, collectively referred to as TMDs, correlate perturbative and non-perturbative physics. Thus the computation of these functions requires sophisticated approaches, typically involving lattice calculations or deriving them through the fitting of experimental data within a comprehensive global QCD analysis. In the case of vacuum TMDs, where there is no cold nuclear medium, there have been significant advancements in the methodologies employed in global analyses of TMDs in the past decade. This progress spans from earlier Gaussian approximations to the current state, where high perturbative accuracy is achieved. This progress is evident in notable works such as~\cite{Anselmino:2013lza,Bacchetta:2017gcc,Scimemi:2017etj,Bertone:2019nxa,Scimemi:2019cmh,Bacchetta:2019sam}. These studies have not only contributed to our understanding of TMD physics but have also propelled the field forward, paving the way for increasingly precise and comprehensive analyses of these crucial correlation functions.

Exploring the femtoscale one-dimensional structure of bound nucleons unveils an array of fascinating challenges in nuclear physics. In scattering experiments involving nuclei, the nuclear medium's influence extends to both partonic and non-perturbative physics, introducing compelling complexities. Studies employing a higher-twist factorization formalism~\cite{Liang:2008vz} and a dipole model~\cite{Mueller:2016gko, Mueller:2016xoc} have elucidated partonic correlations, addressing multiple QCD scattering within the nuclear medium. Recently the connection between these two formalisms has been clarified~\cite{Fu:2023jqv}, leading to a unified picture. 
Furthermore, the modification of the non-perturbative structure of collinear nuclear parton distribution functions (nPDFs) emerges as a potent strategy for accurately describing experimental data. This approach, detailed in a comprehensive review~\cite{Ethier:2020way}, involves the assumption that the perturbative physics remains unchanged while the non-perturbative physics is altered by a nuanced adjustment of the initial non-perturbative parameterization of nPDFs within the collinear factorization formalism, necessitating global analyses of pertinent world data~\cite{Collins:1989gx}. Substantial strides in this field have been achieved~\cite{Eskola:1998df, deFlorian:2003qf, Hirai:2007sx, Eskola:2007my, Schienbein:2009kk, AtashbarTehrani:2012xh, Khanpour:2016pph, Eskola:2016oht, Walt:2019slu, Kovarik:2015cma, AbdulKhalek:2019mzd, AbdulKhalek:2020yuc}, with recent advancements leveraging charged current interactions for flavor tagging. Noteworthy examples include EPPS16~\cite{Eskola:2016oht}, nCTEQ15~\cite{Kova_k_2016}, and nNNPDF~\cite{Khalek_2019}. See also a recent work to study nuclear modification in $e+A$ collisions in the event generator -- eHIJING~\cite{Ke:2023xeo}. These endeavors collectively contribute to unraveling the intricate interplay of partonic and non-perturbative phenomena in the femtoscale structure of bound nucleons.

The ongoing research in this field has also seen significant advancements in unraveling the intricate interplay between perturbative and non-perturbative correlations within QCD matter and collinear fragmentation functions (FFs). In the context of perturbative structure, studies have revealed that the interaction of the struck parton with the QCD medium leads to alterations in the DGLAP evolution of FFs, as exemplified in studies such as \cite{Deng:2009ncl, Ovanesyan:2011xy, Ovanesyan:2011kn,Kang:2014xsa,Chien:2015vja,Ke:2023ixa}. In the context of non-perturbative structures, attention has turned to modifying non-perturbative parameterization of the nuclear fragmentation functions (nFFs). Notable contributions to this avenue of research can be found in works like \cite{Sassot:2009sh, Zurita:2021kli} where the assumption was once again that the perturbative physics has remained unaltered. These investigations shed light on the nuanced dynamics governing the interplay between QCD matter and FFs, offering valuable insights into both perturbative corrections and the non-perturbative aspects of the field.

In our earlier investigation \cite{Alrashed:2021csd}, we introduced a novel approach to incorporate QCD medium contributions into Transverse Momentum Dependent (TMD) measurements by assuming that the perturbative physics remained unaltered while modifying the non-perturbative parameterization of TMDs, leading to the definition of non-perturbative Nuclear Transverse Momentum Distributions (nTMDs). This study marked the first global extraction of nTMDs, leveraging both Semi-Inclusive DIS (SIDIS) experimental data from HERMES~\cite{Airapetian:2007vu} as well as experimental data sets from Drell-Yan production at Fermilab~\cite{Alde:1990im, Vasilev:1999fa}, RHIC~\cite{Leung:2018tql}, and the LHC~\cite{Khachatryan:2015pzs,Aad:2015gta}. The results demonstrated the effectiveness of this framework in describing a comprehensive set of experimental data. Subsequent to our initial study, independent verification was conducted in \cite{Barry:2023qqh}, and further investigations into factorization and resummation were explored in processes involving jets \cite{Gao:2023ulg,Fang:2023thw}.

Following the initial release of our previous study, Jefferson Lab released experimental data sets of the SIDIS multiplicity ratio in \cite{CLAS:2021jhm}. The introduction of this data set grants us the ability to improve our previous analysis in two ways. Firstly, by incorporating these new experimental data into our analysis, we enhance the constraining power of the extraction of the nTMDs. Secondly, this additional constraining power allows us to eliminate our reliance on the LIKEn parameterization of the nFF from \cite{Zurita:2021kli}, providing us with greater flexibility in determining the functional form of the nFF. Moreover, we delve into the impact of the Jefferson Lab measurements in Secs.~\ref{subsec:Fita} and \ref{subsec:Fitb} by conducting two distinct fits: one excluding the Jefferson Lab data, and one incorporating it. This analysis reveals that achieving a simultaneous description of the HERMES and Jefferson Lab data sets requires the introduction of a parameter which controls the non-perturbative evolution of the nTMDs. This finding underscores the additional complexity inherent in TMD measurements involving cold nuclear matter. We present the refined extraction of the nTMDs, and in Sec.~\ref{sec:Preds} we highlight the importance of additional experimental measurements at collider facilities in advancing our understanding of nTMDs by providing predictions and discussions on the constraining power of additional data. 

The paper is organized as follows: In Sections~\ref{sec:Factorization-DIS} and \ref{sec:Factorization-DY}, we provide the factorization and resummation formalism for our analysis. In Sec.~\ref{sec:Num}, we provide details of the numerical treatment of our global analyses. In Sections~\ref{subsec:Fita} and \ref{subsec:Fitb} we provide the results for our fits without and with the Jefferson Lab data. We provide predictions for measurements at Jefferson Lab, and the EICs in Sec.~\ref{sec:Preds}. We conclude in Sec.~\ref{sec:Conclusions}.

\section{Factorization and resummation in DIS}\label{sec:Factorization-DIS}
HERMES~\cite{Airapetian:2007vu} and Jefferson Lab~\cite{CLAS:2021jhm} measured the multiplicity ratio of the hadron $h$
\begin{align}
    R_A^h\left(x,z,P_{h\perp}\right) = \frac{M_A^h\left(x,z,P_{h\perp}\right)}{M_D^h\left(x,z,P_{h\perp}\right)}\,,
\end{align}
where $M_{A/D}^h$ denotes the multiplicity of a nuclear or deuteron target. This multiplicity is defined as the ratio of the SIDIS and DIS cross sections
\begin{align}
    M_{A/D}^h\left(x,z,P_{h\perp}\right) = 
    \dfrac{\dd\sigma_{A/D}^h}{\dd{\mathcal{PS}_{\rm DIS}} \,\dd{z}\,\dd[2]{\boldsymbol{P}_{h \perp}}}/\dfrac{\dd\sigma_{A/D}}{\dd{\mathcal{PS}_{\rm DIS}}}\,
\end{align}
where the numerator is the cross section for SIDIS
\begin{align}
    e\left(\ell \right) + N\left(P\right) \rightarrow e\left(\ell' \right)+ h\left(P_h\right)+X\,, 
\end{align}
while the denominator is the cross section for inclusive DIS
\begin{align}
    e\left(\ell \right) + N\left(P\right) \rightarrow e\left(\ell' \right)+X\,.
\end{align}
In our notation, $e$ denotes the electrons, $N$ denotes the initial state nucleus, $h$ denotes the final-state hadron, and we represent the unobserved state as $X$. For simplicity, we will present the factorization formalisms in the Breit frame, where the momenta of the incoming proton and photon are given by
\begin{align}
    P^\mu = \frac{Q}{x}\frac{n^\mu}{2}+\mathcal{O}\left(\Lambda_{\rm QCD}^2\right)\,,
    \qquad
    q^\mu = -Q\frac{n^\mu}{2}+Q\frac{\bar{n}^\mu}{2}\,,
\end{align}
where the light-cone direction is defined as the direction of the incoming nucleus in the limit that $Q/x\rightarrow \infty$. We have followed the convention to define the light-cone coordinates as $n^\mu = \left(1,0,0,1\right)$ and $\bar{n}^\mu = \left(1,0,0,-1\right)$ in Minkowskian space-time coordinates while the light-cone coordinates as $n\cdot \bar{n} = 2$, which have been widely used in recent studies of soft-collinear effective theory (SCET)~\cite{Bauer:2000ew, Bauer:2000yr, Bauer:2001ct, Bauer:2001yt, Bauer:2002nz}. 

In the SIDIS cross section, $\dd{z}$ and $\dd[2]{\boldsymbol{P}_{h \perp}}$ denote the differential phase space in the hadronic momentum fraction $z$ and transverse momentum of the $h$ relative to the $\gamma^*$-$P$ axis while $\dd{\mathcal{PS}_{\rm DIS}}$ denotes the differential phase space for the inclusive DIS cross section, which is given by $\dd{\mathcal{PS}_{\rm DIS}} = \dd{x} \dd{y}$. The variables $x$, $y$, and $z$ are the parton fraction variables, which are defined in terms of Lorentz invariant scalar products as
\begin{align}
    x = \frac{Q^2}{2 P\cdot q}\,,
    \qquad
    y = \frac{P\cdot q}{P\cdot l}\,,
    \qquad
    z = \frac{P\cdot P_h}{P\cdot q}\,,
\end{align}
while the invariant mass of the incoming space-like photon is given by $Q^2 = -q^2 = -(\ell-\ell')^2$. 
\subsection{DIS}
The expression for the DIS cross section in e-p collisions is well-known and is given by
\begin{align}
\frac{d\sigma}{d\mathcal{PS}}  = \sigma_0^{\rm DIS} \frac{y}{x Q^2} \left[F_2\left(x,Q\right)-\frac{y^2}{1+(1-y)^2}F_L\left(x,Q\right)\right]\,,
\qquad
\sigma_0^{\rm DIS} = \frac{2\pi \alpha_{\rm em}^2}{Q^2}\frac{1+(1-y)^2}{y}
\end{align}
where the $F_2$ and $F_L$ are the unpolarized and longitudinally polarized DIS structure functions. These structure functions can be related to PDFs through an Operator Product Expansion (OPE) as \cite{Bertone:2013vaa}
\begin{align}
    F_J\left(x,Q\right) =   \bqty{\mathcal{C}_J\,\hat{\otimes}\, f}_{q/D}\left(x,Q\right)\,,
\end{align}
where $\mathcal{C}_J$ are the perturbative matching coefficients, $J\in \{2,L\}$, $f$ denotes the vacuum PDF, and the $\hat{\otimes}$ represents the collinear convolutional integral, which is given by
\begin{align}
    \bqty{\mathcal{C}_J \,\hat{\otimes}\, f}_{q/D} \pqty{x,Q}
=
\sum _i \int _x^1 \frac{\dd{\hat{x}}}{\hat{x}} \mathcal{C}_{J,q/i} \pqty{\frac{x}{\hat{x}}, Q}\, f_{i/D}(\hat{x},Q)\,.
\end{align}
In this study, we treat the cross section for e-A collisions as
\begin{align}
\frac{d\sigma_A}{d\mathcal{PS}} = \sigma_0^{\rm DIS}\left[F_2^A\left(x,Q\right)-\frac{y^2}{1+(1-y)^2}F_L^A\left(x,Q\right)\right]\,,
\end{align}
where the nuclear-modified matching coefficients can be written in terms of the nPDFs as
\begin{align}
    F_J^A\left(x,Q\right) = \left[\mathcal{C}_J\,\hat{\otimes}\,f\right]_{a/A}\left(x,Q\right)\,,
\end{align}
where $f$ denotes the nuclear-modified PDF (nPDF). 
\subsection{SIDIS}
\begin{figure}
    \centering
    \includegraphics[height = 2.5in]{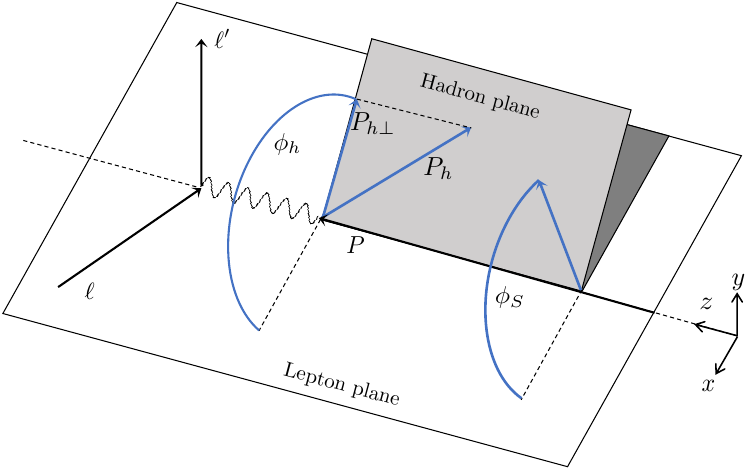}
    \caption{The kinematics for SIDIS in the Breit frame.}
    \label{fig:kinematics-DIS}
\end{figure}

The unpolarized differential cross section for SIDIS is written most conveniently in terms of the unpolarized structure function \cite{Bacchetta:2006tn,Boussarie:2023izj}
\begin{align}\label{eq:sigma-DIS}
\frac{\dd{\sigma}_{N}^h}{\dd{\mathcal{PS}_{\rm DIS}} \,\dd{z}\,\dd[2]{\boldsymbol{P}_{h \perp}}} = \sigma _0\,F_{\rm UU\, N}^h\left(x,z,\bm{P}_{h\perp}\right)+...\,,
\end{align}
where $N$ is either a deuteron or a heavy nucleus and the ellipsis denotes configurations associated with spin dynamics, which we will drop. The structure function for electron-deuteron collisions involves a transverse convolution of the vacuum TMD PDF and the vacuum TMD FF, which is given explicitly as \cite{Ji:2004wu}
\begin{align}
    F_{\rm UU\, D}^h\left(x,z,\bm{P}_{h\perp}\right) = H_{\rm DIS}\pqty{Q,\mu}  \sum_q e_q^2 \int & \dd[2]{\boldsymbol{k}_{\perp}} \dd[2]{\boldsymbol{p}_{\perp}}\, \delta^2\pqty{z\bm{k}_\perp+\bm{p}_\perp -\bm{P}_{h\perp}} \nn \\ 
    & \times f_{1\,q/D}\pqty{x,k_\perp,\mu,\zeta _1}\, D_{1\,h/q}\pqty{z,p_\perp,\mu,\zeta _2}\,,
\end{align}
where $H_{\rm DIS}$ is the hard function for this process that is given at one loop in the appendix in Eq.~\eqref{eq:hard-DIS}. Additionally, we have introduced the scales $\mu$, $\zeta_1$, and $\zeta_2$, which are the renormalization scale and the Collins-Soper scales of the TMDs. Note that while the TMDs depend on the scales $\zeta_1$ and $\zeta_2$, the cross section depends only on the product of these two scales, which is given by $\zeta_1 \zeta_2 = Q^4$. Lastly, we note that the transverse convolution integrates over $\bm{k}_\perp$ and $\bm{p}_\perp$, the transverse momenta of the incoming quark with respect to the incoming hadron and the transverse momentum of the final-state hadron with respect to the fragmenting quark. This transverse convolution integrals is simplified by working in $b$-space, the conjugate space to $-\bm{P}_{h\perp}/z$, to
\begin{align}
    F_{\rm UU\, D}^h\left(x,z,\bm{P}_{h\perp}\right) = H_{\rm DIS}\pqty{Q,\mu} \sum_q e_q^2 & \int \frac{b \dd{b}}{2\pi} J_0 \pqty{\frac{b\, P_{h\perp}}{z}} \\
    &\times f_{1\, q/D}\pqty{x,b,\mu,\zeta_1}\, D_{1\, h/q}\pqty{z,b,\mu,\zeta_2}\,, \nn
\end{align}
where $J_0$ is the zeroth Bessel function of the first kind. In this expression, we have introduced the $b$-dependent TMDs, which are defined as
\begin{align}
f_{1\, q/D} (x, b,\mu, \zeta)
& =
\int \dd[2]{\boldsymbol{k}_{\perp}} e^{-i \boldsymbol{b} \cdot \boldsymbol{k}_\perp} f_{1\, q/D} \pqty{x,k_\perp,\mu,\zeta}\,,
\\
D_{1\, h/q} (z, b,\mu, \zeta)
& =
\frac{1}{z^2} \int \dd[2]{\boldsymbol{p}_{\perp}} e^{-i \boldsymbol{b} \cdot \boldsymbol{p}_{\perp} / z} D_{1\, h/q} \pqty{z, p_{\perp},\mu, \zeta}\,.
\end{align}
At this point, we note that in the hard function and the TMDs, there exist large logarithms that must be resummed. The resummation of these logarithms is performed by solving the evolution equations associated with each contribution. The TMD PDF and TMD FF depend on both the renormalization scale $\mu$ and the rapidity scale $\zeta$ and thus obey a coupled differential equation. The hard function however depends only on the renormalization group scale and thus its evolution is given by
\begin{align}
    \frac{d}{d\ln\mu}\ln H_{\rm DIS}\left(Q,\mu\right) = \gamma_\mu^H(\mu)\,,
\end{align}
where $\gamma_\mu^H$ is the anomalous dimension of the hard function and is given in the appendix in Eq.~\eqref{eq:anom-hard}. The solution of this evolution equation is given by
\begin{align}
    H_{\rm DIS}\left(Q,\mu\right) = H_{\rm DIS}\left(Q,\mu_H\right)\, U\left(\mu_H,\mu\right)\,,
    \qquad
    U_H\left(\mu_H,\mu\right) = \exp\left[\int_{\mu_H}^\mu \frac{d\mu'}{\mu'}\gamma_\mu^H(\mu')\right]\,,
\end{align}
where the term $U_H$ denotes the perturbative evolution of the hard factor and $\mu_H$ denotes an arbitrary initial scale for the hard function and we use the language of the TMD handbook \cite{Boussarie:2023izj}. This scale is usually taken to be $\mu_H = Q$ to eliminate logs in the fixed order expression for the hard function. Here we allow this scale dependence to be general and discuss our choice of $\mu_H$ later in the paper. The coupled RG and Collins-Soper evolution equations \cite{Collins:1981uk} which govern the TMDs are given by
\begin{align}
    \frac{d}{d\ln\mu}\ln F(w,b,\mu,\zeta) = \gamma_\mu\left(\mu,\zeta\right)\,,
    \qquad
    \frac{d}{d\ln\zeta}\ln F(w,b,\mu,\zeta) = \gamma_\zeta\left(b,\mu\right)\,,
\end{align}
where $\gamma_\mu$ and $\gamma_\zeta$ are the anomalous dimension and the rapidity anomalous dimension of the TMDs. Additionally, $F\in \left\{f_1,D_1\right\}$ and $w \in \left\{x,z\right\}$ and we note that the anomalous dimensions ($\gamma_\mu$ and $\gamma_\zeta$) of the $f_1$ and $D_1$ are the same and are given in the appendix at NNLL in Eqs.~\eqref{eq:anom-f}, \eqref{eq:anom-D}, and \eqref{eq:anom-zeta}. The solution of this evolution equation is
\begin{align}\label{eq:TMD PDF}
    F\left(w,b,\mu,\zeta\right) = F\left(w,b,\mu_i,\zeta_i\right)\, U\left(\mu_i, \mu;\zeta\right)\, Z\left(b,\zeta_i, \zeta;\mu_i\right)\,,
\end{align}
where the perturbative evolution is governed by the Sudakov terms
\begin{align}
    U\left(\mu_i,\mu;\zeta\right) = \exp\left[\int_{\mu_i}^{\mu} \frac{d\mu'}{\mu'}\gamma_\mu\left(\mu',\zeta\right)\right]\,, \qquad
    Z\left(b,\mu_i,\mu;\zeta\right) = \left(\frac{\zeta}{\zeta_i}\right)^{\gamma_\zeta\left(b,\mu_i\right)}\,.
\end{align}
In these expressions, we have introduced the natural scales for initial TMDs, $\mu_i$ and $\zeta_i$, which should be taken to eliminate large logs in the fixed order expressions for the TMDs. As in the case of the hard function, we defer discussion on these scales.

The final expression for the cross section can be obtained by noting that the vacuum TMDs can be perturbatively matched onto collinear distributions via an OPE in the limit of small $b$ as \cite{Collins:2011zzd}
\begin{align}
f_{1\, q/D} \pqty{x, b,\mu, \zeta} & = \bqty{C \otimes f}_{q/D} \pqty{x, b,\mu, \zeta}\,, \\
D_{1\, h/q} \pqty{z, b,\mu, \zeta} & = \frac{1}{z^2} \bqty{\hat{C}\otimes D}_{h/q} \pqty{z, b,\mu, \zeta}\,, \nn 
\end{align}
where $f$ and $D$ on the right-hand sides of these expressions denote the collinear PDF and FF while the $C$ and $\hat{C}$ functions denote the matching coefficients which we provide up to one loop in the appendix. In this expression, we have used the short-hand notation for the collinear convolutions
\begin{align}
\bqty{C \otimes f}_{q/D} \pqty{x, b,\mu, \zeta}
=
\sum _i \int _x^1 \frac{\dd{\hat{x}}}{\hat{x}} C_{q/i} \pqty{\frac{x}{\hat{x}}, b,\mu, \zeta}\, f_{i/D}(\hat{x},\mu)\,,
\\
\bqty{\hat{C} \otimes D}_{h/q} \pqty{z, b,\mu, \zeta}
=
\sum _i \int _z^1 \frac{\dd{\hat{z}}}{\hat{z}} D_{h/i}(\hat{z},\mu)\, \hat{C}_{i/q} \pqty{\frac{z}{\hat{z}}, b,\mu, \zeta} \,.
\end{align}
By studying the one loop expressions for the matching functions in Eq.~\eqref{eq:match}, we see that the logarithms are minimized by taking the initial scale choice that $\mu_i = \sqrt{\zeta_i} = \mu_b$ where $ \mu_b = 2 e^{-\gamma_E}/b$ denotes the so-called `natural scale' for the TMDs. After taking this into consideration, the expressions for the matched TMDs are given by
\begin{align}\label{eq:match-PDF}
f_{1\, q/D} \pqty{x, b,\mu, \zeta} & = \bqty{C \otimes f} \pqty{x, b,\mu_i, \zeta_i}\, U\left(\mu_i,\mu;\zeta\right)\, Z\left(b,\zeta_i,\zeta;\mu_i\right)\, U_{\rm NP}^{f^A}\left(x,b,\zeta\right)\,, \\
\label{eq:match-FF}
D_{1\, h/q} \pqty{z, b,\mu, \zeta} & = \frac{1}{z^2} \bqty{\hat{C}\otimes D} \pqty{z, b,\mu_i, \zeta_i} \, U\left(\mu_i,\mu;\zeta\right)\, Z\left(b,\zeta_i,\zeta;\mu_i\right)\, U_{\rm NP}^D\left(z,b,\zeta\right)\,.
\end{align}
In these expressions, $U_{\rm NP}$ denote the non-perturbative Sudakov terms for the TMD PDF and TMD FF that will be discussed in Sec.~\ref{sec:Num}. After taking into account the matching, the final expression for the structure function is given by
\begin{align}\label{eq:fac-DIS}
    F_{\rm UU\, D}^h & \left(x,z,\bm{P}_{h\perp}\right) = H_{\rm DIS}\pqty{Q,\mu_H} \frac{1}{z^2}  \sum_q e_q^2 \int \frac{b \dd{b}}{2\pi} J_0 \pqty{\frac{b\, P_{h\perp}}{z}} \, U_{\rm tot}\left(b,\mu_i,\mu_H,\zeta_i,Q^2\right) \\
    & \hspace{1cm} \times \bqty{\hat{C}\otimes D}_{h/q} \pqty{z, b,\mu_i, \zeta_i} \, \bqty{C \otimes f}_{q/D} \pqty{x, b,\mu_i, \zeta_i}\, U_{\rm NP}^f(x,b,\zeta_f)\, U_{\rm NP}^D(z,b,\zeta_D) \,, \nn
\end{align}
where the Sudakov for the cross section is given by
\begin{align}
    U_{\rm tot}\left(b,\mu_i,\mu_H,\zeta_i,Q^2\right) = U_H\left(\mu_i,\mu_H;Q^2\right)\, Z^2\left(b,\zeta_i, Q^2;\mu_i\right)\,,
\end{align}
where we have used the relation that $\zeta_1 \zeta_2 = Q^4$ and taken the TMDs to initialize at the same $\mu$ and $\zeta$ scales. In this expression, we see that the scale $\mu_H$ is still free. For phenomenology, we will take the canonical scale choice $\mu_H = Q$, which eliminates logs in the hard function. 

In our analysis, we write the structure function for electron-nucleus collisions as
 \begin{align}
    F_{\rm UU\, A}^h\left(x,z,\bm{P}_{h\perp}\right) = H_{\rm DIS}\pqty{Q,\mu}  \sum_q e_q^2 \int & \dd[2]{\boldsymbol{k}_{\perp}} \dd[2]{\boldsymbol{p}_{\perp}}\, \delta^2\pqty{z\bm{k}_\perp+\bm{p}_\perp -\bm{P}_{h\perp}} \nn \\ 
    & \times  D_{1\,h/q}^A\pqty{z,p_\perp,\mu,\zeta _2}\, f_{1\,q/A}\pqty{x,k_\perp,\mu,\zeta _1}\,.
\end{align}
To reach this expression, we have assumed that the nuclear medium acts to modify the TMDs while leaving the hard physics unchanged. We note that in our treatment both the TMD PDF and the TMD FF are modified due to the medium. These nTMDs can be matched onto collinear distributions in a manner analogous to the vacuum TMDs. Namely, we write
\begin{align}\label{eq:match-A}
f_{1\, q/A} \pqty{x, b,\mu, \zeta} & = \bqty{C \otimes f}_{q/A} \pqty{x, b,\mu_i, \zeta_i}\, U\left(\mu_i,\mu;\zeta\right)\, Z\left(b,\zeta_i,\zeta;\mu_i\right)\, U_{\rm NP}^{f^A}\left(x,b,\zeta,A\right)\,, \\
\qquad
D_{1\, h/q}^A \pqty{z, b,\mu, \zeta} & = \frac{1}{z^2} \bqty{\hat{C}\otimes D^A}_{h/q}\pqty{z, b,\mu_i, \zeta_i} \, U\left(\mu_i,\mu;\zeta\right)\, Z\left(b,\zeta_i,\zeta;\mu_i\right)\, U_{\rm NP}^{D^A}\left(z,b,\zeta,A\right)\,.\nn
\end{align}
In this expression, we use $D^A$ to denote the nuclear modified FF. Additionally, we have introduced the non-perturbative Sudakov terms for the nTMDs. Lastly, we note to arrive at this expression, we have left the perturbative evolution factors $U$ and $Z$ the same as that of the vacuum TMDs. Following these assumptions, the final expression for the structure functions in e-A collisions is given by
\begin{align}\label{eq:fac-DIS-A}
    & F_{\rm UU\, A}^h \left(x,z,\bm{P}_{h\perp}\right) = H_{\rm DIS}\pqty{Q,\mu_H} \frac{1}{z^2}  \sum_q e_q^2 \int \frac{b \dd{b}}{2\pi} J_0 \pqty{\frac{b\, P_{h\perp}}{z}} \, U_{\rm tot}\left(b,\mu_i,\mu_H,\zeta_i,Q^2\right) \nn \\
    & \bqty{\hat{C}\otimes D^A}_{h/q} \pqty{z, b,\mu_i, \zeta_i} \,\bqty{C \otimes f}_{q/A} \pqty{x, b,\mu_i, \zeta_i}\, U_{\rm NP}^{f^A}(x,b,\zeta_f,A)\, U_{\rm NP}^{D^A}(z,b,\zeta_D,A) \,.
\end{align}

\section{Factorization and resummation in Drell-Yan}\label{sec:Factorization-DY}
\begin{figure}
    \centering
    \includegraphics[height = 2.5in]{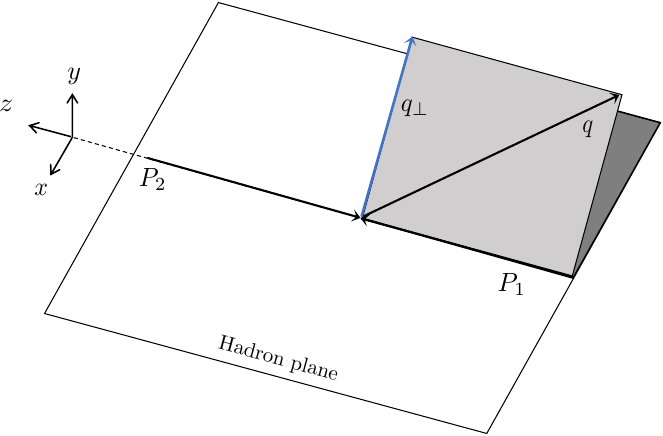}
    \caption{The kinematics for Drell-Yan in the hadronic CM frame. Here $P_2$ represents the momentum of the nucleus.}
    \label{fig:kinematics-DY}
\end{figure}

We begin this section by labeling the momenta of the external particles as
\begin{align}
    p(P_1)+N(P_2) \rightarrow \gamma^*/Z(q)+X \rightarrow l^+(\ell)+l^-(\ell')+X
\end{align}
where $p$ denotes the incoming proton, $N$ is once again an incoming nucleus, $\gamma^*/Z$ denotes the produced neutral vector boson, while $l^{\pm}$ denote the positively (negatively) charged leptons. 
For this process, we define the light-cone directions in terms of the incoming nucleons
\begin{align}
    P_1^\mu = \sqrt{S}\frac{n^\mu}{2}+\mathcal{O}\left(\Lambda_{\rm QCD}^2\right)\,,
    \qquad
    P_2^\mu = \sqrt{S}\frac{\bar{n}^\mu}{2}+\mathcal{O}\left(\Lambda_{\rm QCD}^2\right)\,,
\end{align}
where we have chosen to work in the nucleon-nucleon center of mass frame (CM). In this frame, the produced vector boson can be parameterized in terms of its mass, transverse momentum, and rapidity as
\begin{align}
    q^\mu = \left(M_\perp e^y,M_\perp e^{-y},\bm{q}_\perp\right)\,,
\end{align}
where 
\begin{align}
    Q^2 = q^2\,,
    \qquad
    y = \frac{1}{2}\ln\left(\frac{\bar{n}\cdot q}{n\cdot q}\right)\,,
\end{align}
and $M_\perp = \sqrt{Q^2+q_\perp^2}$ is the transverse mass of the vector boson. The Bjorken fractions can be defined in terms of the external momenta as
\begin{align}
    x_1 = \frac{Q^2}{2 P_1\cdot q} = \frac{Q}{\sqrt{S}}e^y\,,
    \qquad
    x_2 = \frac{Q^2}{2 P_2\cdot q} = \frac{Q}{\sqrt{S}}e^{-y}\,,
\end{align}
while the Feynman variable is given by $x_F = x_1-x_2$. The differential cross section for p-p collisions for this process can be written as
\begin{align}
    \frac{d\sigma}{dy\,d^2q_\perp\,dQ^2} = \sigma_0^{\rm DY} W_{\rm UU}\left(y,q_\perp,Q\right)\,,
    \qquad
    \sigma_0^{\rm DY} = \frac{4\pi \alpha^2}{3 N_c Q^2 S}\,,
\end{align}
where the structure function can be written as transverse convolution of the quark and anti-quark TMD PDFs as \cite{Collins:1984kg,Bacchetta:2019sam}
\begin{align}
    W_{\rm UU}(y, q_\perp, Q) = H_{\rm DY}\left(Q,\mu\right) & \mathcal{P}\left(y,q_\perp,Q\right)\sum_q c_q\left(Q^2\right) \int \dd[2]{\bm{k}_{1\perp}} \dd[2]{\bm{k}_{2\perp}} \delta^2\left(\bm{q}_\perp-\bm{k}_{1\perp}-\bm{k}_{2\perp}\right) \nn\\
    & f_{1\, q/p}\left(x_2,k_{1\perp},\mu,\zeta_1\right) f_{1\, \bar{q}/N}\left(x_2,k_{2\perp},\mu,\zeta_2\right) \,.
\end{align}
In this expression $H_{\rm DY}$ denotes the hard function for Drell-Yan, which is given at one loop in the appendix in Eq.~\eqref{eq:hard-DY}. Due to the interference between the $\gamma^*$ and $Z$ bosons, the couplings $c_q$ take on a complicated form and are given by \cite{Bacchetta:2019sam}
\begin{align}
    c_q\left(Q\right) = e_q^2-2 e_q V_q V_l \chi_1\left(Q^2\right)+\left(V_l^2+A_l^2\right)\left(V_q^2+A_q^2\right)\chi_2\left(Q^2\right)\,,
\end{align}
\begin{align}
    \chi_1\left(Q^2\right) &= \frac{1}{4\sin^2 \theta_W \cos^2 \theta_W}\frac{Q^2(Q^2-M_Z^2)}{(Q^2-M_Z^2)^2+M_Z^2 \Gamma_Z^2}\,,
    \\
    \chi_2\left(Q^2\right) &= \frac{1}{16\sin^4 \theta_W \cos^4 \theta_W}\frac{Q^4}{(Q^2-M_Z^2)^2+M_Z^2 \Gamma_Z^2}\,.
\end{align}
In these expressions, $e_q$, $V_q$, $A_q$ represent the electric, vector, and axial charges of the quark of flavor $q$. Additionally, $V_l$ and $A_l$ are the vector and axial charges of the lepton. Lastly, $\theta_W$ is the weak mixing angle while $M_Z$ and $\Gamma_Z$ are the mass and widths of the $Z$ boson. 

In the expression for the cross section, we have also introduced the fiducial cut $\mathcal{P}$, which characterizes experimental cuts on the detector associated with the final-state lepton pair. The exact expression for the fiducial cuts is given by
\begin{align}
    \mathcal{P}\left(y,q_\perp,Q\right) = \int_{\ell\, \ell'}\, g^{\mu\nu}_\perp L_{\mu\nu}\left(\ell, \ell'\right)\, \Theta(\ell,\ell')/\int_{\ell\, \ell'}\, g^{\mu\nu}_\perp L_{\mu\nu}\left(\ell, \ell'\right)\,
\end{align}
where we use the short-hand that
\begin{align}
    \int_{\ell\, \ell'}  = \int d^4\ell\, d^4 \ell'\, \delta^4\left(q-\ell-\ell'\right)\,.
\end{align}
In the first expression, $L_{\mu\nu}$ is the leptonic tensor and
\begin{align}
    g^{\mu\nu}_\perp = g^{\mu\nu} - \hat{z}^\mu \hat{z}^\nu - \hat{t}^\mu \hat{t}^\nu\,,
\end{align}
where the time direction is given by $\hat{t}^\mu = q^\mu/Q$, while the $z$ direction is given by the three momentum of the incoming hadron's momentum $P_1$. The transverse Minkowski metric projects out the leading power contribution to the cross section \cite{Gamberg:2022lju}. Lastly, the restrictions on the phase space imposed by the fiducial cuts are given in the case of the CMS data \cite{CMS:2015zlj} takes the simple form
\begin{align}
    \Theta\left(\ell,\ell'\right) = \Theta\left(y_{\rm cut}-|y_{\ell}|\right) \Theta\left(y'_{\rm cut}-|y_{\ell'}|\right) \Theta\left(\ell_\perp - \ell_{\perp\, \rm{cut}}\right) \Theta\left(\ell_\perp' - \ell'_{\perp\, \rm{cut}}\right)\,,
\end{align}
where $y_\ell$, $y_\ell'$, $\ell_\perp$, and $\ell_\perp'$ denote the rapidity and the transverse momenta of the final-state leptons. 

The expression for the cross section can once again be simplified by going to $b$-space and is given as the inverse Fourier transform
\begin{align}\label{eq:fac-DY}
    & W_{\rm UU} \left(y, q_\perp, Q\right) = H_{\rm DY}\pqty{Q,\mu_H} \mathcal{P}\left(y,q_\perp, Q\right)\sum_q c_q\left(Q\right)\, \int \frac{b \dd{b}}{2\pi} J_0 \pqty{b\, q_\perp}\, U_{\rm tot}\left(b,\mu_i,\mu_H,\zeta_i,Q^2\right) \nn \\
    & \hspace{0.9cm} \times \bqty{C \otimes f}_{q/p} \pqty{x_1, b,\mu_i, \zeta_i}\, \bqty{C \otimes f}_{\bar{q}/p} \pqty{x_2, b,\mu_i, \zeta_i} \, U_{\rm NP}^f(x_1,b,\zeta_{f_1})\, U_{\rm NP}^f(x_2,b,\zeta_{f_2})\,,
\end{align}
where $\zeta_1 \zeta_2 = Q^4$. Following the discussion in SIDIS, the cross section for p-A collisions can be written as
\begin{align}\label{eq:fac-DY-A}
    & W_{\rm UU\, A} \left(y, q_\perp, Q\right) = H_{\rm DY}\pqty{Q,\mu_H} \mathcal{P}\left(y,q_\perp, Q\right)\sum_q c_q\left(Q\right)\, \int \frac{b \dd{b}}{2\pi} J_0 \pqty{b\, q_\perp}\, U_{\rm tot}\left(b,\mu_i,\mu_H,\zeta_i,Q^2\right) \nn \\
    & \hspace{0.4cm} \times \bqty{C \otimes f}_{q/p} \pqty{x_1, b,\mu_i, \zeta_i}\, \bqty{C \otimes f}_{\bar{q}/A} \pqty{x_2, b,\mu_i, \zeta_i} \, U_{\rm NP}^{f}(x_1,b,\zeta_{f_1})\, U_{\rm NP}^{f^A}(x_2,b,\zeta_f,A)\,,
\end{align}
where we have used the conventions of the LHC data to place the nucleus to be going in the $-z$ direction.

\section{Numerical treatment}\label{sec:Num}
In this paper, we address the question of how the Jefferson Lab data influences the global analysis of the nFFs. For this purpose, we will present two fits, which we denote fit(a) and fit(b). In fit(a), we consider the complete set of Drell-Yan data but consider only the HERMES SIDIS data. In fit(b) we again take the complete set of Drell-Yan data but consider both the HERMES and Jefferson Lab data sets. In the following sections, we will discuss the numerical treatment of each of these fits. 

\subsection{Data selection}\label{subsec:Data}
The HERMES measurement of the multiplicity ratio was obtained by counting events that fell into a particular region of $z$, $P_{h\perp}$, $\nu$, and $Q$. By performing this analysis experimentalists can ``project" the events of their measurement to generate experimental data that depends on a particular kinematic variable by integrating over the other kinematic variables. Thus in the case of the HERMES data, the experimental data can be expressed as being $z$, $P_{h\perp}$, $\nu$, or $Q$ dependent. Because of this process, the experimental data points from one projection are correlated with the data points of another projection. For instance, the experimental data points for the $z$ and $P_{h\perp}$ projections were obtained by binning the same events. If we were to consider the $z$ and $P_{h\perp}$ projections of the experimental data, our fit would be very sensitive to statistical fluctuations in the measurement and this would result in an underestimation of the fit uncertainties. To avoid this issue, for the HERMES data we must then consider only a single projection of the experimental data. As we will later discuss, for fit(a), we will use the $z$ dependent data while for fit(b), we will use the $P_{h\perp}$ dependent data. Additionally, to avoid correlations with projections of the Jefferson Lab data, we will use the $P_{h\perp}$ dependent data in fit(b).

We can now consider removing experimental data based on kinematic arguments. We begin this discussion by noting that in the perturbative expansion of the SIDIS cross section, there are threshold logs of the form $\ln(1-z)$ which become large as $z$ approaches 1. To avoid introducing large non-perturbative contributions to the fit, the consideration of these data requires the resummation of these threshold logarithms. These data can potentially be treated by performing a simultaneous resummation of TMD and threshold logarithms using the formalism in e.g.~\cite{Kang:2022nft}. However, a joint TMD-threshold global analysis has never been performed and is beyond the scope of this paper. To reduce the contributions of the threshold logarithms, we thus impose the constraint that $z<0.7$ in all SIDIS data. In addition, we must also remove all experimental data which falls outside of the TMD region. The TMD region is formally defined in SIDIS and Drell-Yan in the region where the transverse momenta of the partons is much smaller than the hard scale $\Lambda_{\rm QCD}\lesssim k_\perp \ll Q$ and $\Lambda_{\rm QCD}\lesssim p_\perp \ll Q$. For phenomenological purposes, experimental data that is outside of the TMD region is pruned by enforcing the kinematic cuts
\begin{align}
    \frac{P_{h\perp}}{z} \leq \delta_{\rm DIS}\, Q\,,
    \qquad
    q_\perp \leq \delta_{\rm DY}\, Q\,,
\end{align}
where $\delta_{\rm DIS/DY}$ are constants that should be small.

While these kinematic constraints on the experimental data are well justified based on theoretical arguments, these restrictions strongly limit the number of available experimental data. The current set of experimental data for SIDIS tends to be at small $Q$ and large $z$, thus the number of data in the TMD region is severely limited. In the case of the HERMES data, these limitations are particularly drastic. For a cut value of $\delta_{\rm DIS} = 0.3$, we are left with 9 points for the $P_{h\perp}$ projection of the HERMES data and zero points for $z$ projection of this set. For a cut of $\delta_{\rm DIS} = 0.5$, we are left with 18 points for $P_{h\perp}$ projection and zero points for the $z$ projection. For the case of fit(a), which relies solely on the HERMES data, this small number of experimental data points becomes problematic for constraining the functional form of the collinear nFF. The issue of the small number of experimental data when cutting in $P_{h\perp}/z$ has been circumvented in the literature by cutting on $P_{h\perp}$ rather than $P_{h\perp}/z$, see for instance \cite{Bacchetta:2020gko,Echevarria:2020hpy,Alrashed:2021csd}. To circumvent this issue, we choose the cut $P_{h\perp}^2<0.3$ GeV$^2$, which leaves us with 47 points for the $z$ projection and 27 points for the $P_{h\perp}$ projection of the data. The larger number of experimental data for the $z$ projection is an attractive feature. Furthermore, we find that the $P_{h\perp}$ projection of the data covers a very narrow range in $z$, $0.38-0.39$, and thus would serve very weakly in constraining the $z$ dependence of the TMD FF. The $z$ projection of the HERMES data however covers the kinematic region $0.15<z<0.65$. Thus we choose for fit(a) to use the $z$ projection of the experimental data. For fit(b) however, the Jefferson Lab supplements the HERMES data set. Thus for fit(b), we use the $P_{h\perp}$ projection of the HERMES data and choose the more rigorous TMD kinematic cut $\delta_{\rm DIS} = 0.5$ which leaves us with 18 HERMES data points.

\subsection{Parameterization of the vacuum cross section}\label{subsec:vNP}
The non-perturbative contributions to the TMD cross sections enter from the initial parameterization of the collinear distributions and TMDs. The non-perturbative collinear contributions are controlled by the PDF and the FF while the non-perturbative transverse contributions are controlled by the non-perturbative Sudakov factors. In this section, we will discuss the details regarding how we parameterize each of these degrees of freedom. 

In this paper, we use the DEHSS parameterization \cite{deFlorian:2014xna} for the vacuum FF~\footnote{We note that there is a newer parameterization of vacuum FFs for pions~\cite{Borsa:2021ran}.}, in which the authors performed an NLO extraction of the FFs from single-inclusive pion production in electron-positron annihilation (SIA) and collinear SIDIS data. We note that while the time-like evolution kernels for the FFs have been derived at NNLO, the partonic cross sections for SIDIS are currently unknown. As a result, the highest precision simultaneous extractions of the FFs from SIA and SIDIS are currently known only to NLO. This detail alone serves as a bottleneck in the perturbative treatment of global extractions of TMDs, although alternative treatments of this bottleneck have been studied in \cite{Abele:2021nyo,Borsa:2022vvp,AbdulKhalek:2022laj,Bacchetta:2022awv}. Due to this bottleneck, along with additional bottlenecks that are discussed in the next section, in this paper, all non-perturbative parameterizations will be taken from global analyses at NLO. To consistently treat the perturbative accuracy of the PDF and the FF, we use the CT18ANLO parameterization for the collinear PDF from \cite{Hou:2019efy}. As we will discuss in the next section, there are additional considerations for using CT18ANLO. Finally, we will follow the parameterization of Ref.~\cite{Sun:2014dqm,Kang:2015msa,Echevarria:2020hpy,Alrashed:2021csd} for the TMDs. In \cite{Sun:2014dqm}, the authors performed an NLO extraction of the TMDs using the parameterization
\begin{align}
U_{\rm NP}^{f}(x,b,\zeta) & = \exp\left\{-g_q b^2 -\frac{g_2}{2}\ln\left(\frac{b}{b_*}\right)\ln\left(\frac{\zeta}{\zeta_0}\right)\right\}\,, \\
U_{\rm NP}^{D}(z,b,\zeta) & = \exp\left\{-g_h \frac{b^2}{z^2} -\frac{g_2}{2}\ln\left(\frac{b}{b_*}\right)\ln\left(\frac{\zeta}{\zeta_0}\right)\right\}\,.
\end{align}
In these expressions, the logarithmic terms are associated with the non-perturbative contribution of the Collins-Soper kernel in the region of large $b$. In those logarithms, we have introduced logs of the initial TMD scale $\zeta_0 = 2.4$ GeV$^{2}$ and the final Collins-Soper scale. The Gaussian terms in these expressions denote the non-perturbative widths of the TMDs in $b$-space at the scale $\sqrt{\zeta} = \sqrt{\zeta}_0$. The values of the parameters are given by $g_{2}$ = 0.84, $g_{h}$ = 0.042 GeV$^{2}$, and $g_{q}$ = 0.106 GeV$^{2}$.

In this paper, all fixed order terms are taken at NLO accuracy. Additionally, the perturbative Sudakov term involves resumming logarithms of the form $\alpha_s^n \ln^m\left(q_\perp/\mu_H\right)$. These terms can be organized by their magnitude as
\begin{align}
    \rm{LL} \sim \sum_{n= 0}^\infty \alpha_s^n \ln^{n+1}\left(\frac{\mu_H}{q_\perp}\right)\,, \qquad
    \rm{NLL} \sim \sum_{n= 0}^\infty \alpha_s^n \ln^{n}\left(\frac{\mu_H}{q_\perp}\right)\,, \qquad
    \rm{NNLL} \sim \sum_{n= 0}^\infty \alpha_s^n \ln^{n-1}\left(\frac{\mu_H}{q_\perp}\right)\,.
\end{align}
Here there are additional higher logarithmic terms that are not considered. As the TMD region is defined at $q_\perp \ll Q$, these logarithms become large enough that they scale like $\ln\left(\mu_H/q_\perp\right) \sim 1/\alpha_s$ so that the NNLL terms are of the same order as the fixed order contributions. Thus in this paper, we work at NLO+NNLL accuracy. 

In Eqs.~\eqref{eq:fac-DIS} and \eqref{eq:fac-DY}, we saw that the evolution equations played the role of evolving the cross section from the natural scales of the TMDs $\mu_b = 2 e^{-\gamma_E}/b$ up to the hard scale of the process $\mu_H = Q$. These expressions involved integration over all values of $b$. However, at large $b$, the scale entering into the perturbative evolution can become non-perturbative $\mu_b \sim \Lambda \rm_{QCD}$. Thus the scale $\mu_b$ needs to be parameterized to extrapolate smoothly between the perturbative small-$b$ region and the non-perturbative large-$b$ region. This has been extensively studied for instance in  using the $b_*$-prescription \cite{Collins:2014jpa,Aidala:2014hva,Sun:2014dqm,Landry:2002ix,Konychev:2005iy,Bacchetta:2017gcc,Bacchetta:2022awv}. In this work, we follow the standard $b_*$-prescription where
\begin{align}\label{eq:bstar}
  b_* \equiv b /\sqrt{1+b^2/b_{\rm max}^2}\,,
\end{align}
as in \cite{Collins:1984kg}, where we choose $b_{\rm max}=1.5$ GeV$^{-1}$ to tame the large $b$ behavior. Namely in the small $b$ and large $b$ regions, the $b_*$ prescription has the behavior
\begin{align}
    b_* = b\left(1+\frac{1}{2}\frac{b^2}{b_{\rm max}^2}\right)+\mathcal{O}\left(\frac{b^4}{b_{\rm max}^4}\right)\,,
    \qquad
    b_* = b_{\rm max}\left(1-\frac{1}{2}\frac{b_{\rm max}^2}{b^2}\right) +\mathcal{O}\left(\frac{b_{\rm max}^4}{b^4}\right)\,.
\end{align}
Thus in our paper, we take the initial scale choices $\mu_i = \sqrt{\zeta_i} = 2e^{-\gamma_E}/b_*$. From the outlined asymptotic behavior of the $b_*$ prescription, this scale serves as a small modification to the perturbative physics while avoiding the complications associated with the un-starred prescription near $\mu_b \sim \Lambda_{\rm QCD}$. Lastly, to perform the Fourier transforms in the expressions for the cross sections, we use the FBT code from \cite{Kang:2019ctl}.

\subsection{Parameterization for the nTMDs}\label{subsec:nNP}
Analogous to the TMDs, the nTMDs contain non-perturbative contributions from both the collinear distributions and the non-perturbative Sudakov terms. While the nPDF has been studied extensively, obtaining the nFF is an objective of this study. In this section, we begin by discussing our parameterization of the nFF and then move on to discuss our parameterization of the Sudakov terms.

To parameterize the non-perturbative Sudakovs of the nuclei, we modify the parameterization of our previous study
\begin{align}
U_{\rm NP}^{f^A}(x,b,\zeta) & = U_{\rm NP}^{f}(x,b,\zeta) \exp\left\{-g_q^A\,\left(A^{1/3}-1\right)\,b^2 \left(\frac{\zeta_A}{\zeta}\right)^\Gamma\right\}\,, \\
U_{\rm NP}^{D^A}(x,b,\zeta) & = U_{\rm NP}^{D}(x,b,\zeta) \exp\left\{-g_h^A\,\left(A^{1/3}-1\right)\,\frac{b^2}{z^2} \left(\frac{\zeta_A}{\zeta}\right)^\Gamma\right\}\,,
\end{align}
where $\zeta_A = 1$ GeV$^2$ while $g_q^A$ and $g_h^A$ represent modifications to the width due to the nuclear medium. The logarithm represents modifications associated with possible evolution effects associated with emissions in the nuclear medium. We note that for $\Gamma = 0$, this parameterization returns to that of our previous paper.

To parameterize the nPDF, we use the state-of-the-art EPPS21 parameterization \cite{Eskola:2021nhw} where the nPDFs are written in terms of the vacuum PDFs as
\begin{align}
    f_{i/A}\left(x,Q\right) = R_{i}^A\left(x,Q\right)\,f_{i/p}\left(x,Q\right)\,,
\end{align}
where the ratios $R$ are provided by EPPS21. To parameterize the collinear PDFs, the EPPS21 analysis used the CT18ANLO parameterization. This reason is why we chose to use the CT18ANLO parameterization in the previous section. 

To discuss the parameterization of the nFF, it is useful to note that the DEHSS study parameterized the FFs at the scale $\mu_0=1\,\mathrm{GeV}$ as a normalized polynomial
\begin{align}
\label{eq:ff-input}
D_i^{\pi^+}\!(z,\mu_0) =
\frac{N_i z^{\alpha_i}(1-z)^{\beta_i} [1+\gamma_i (1-z)^{\delta_i}] }
{B[2+\alpha_i,\beta_i+1]+\gamma_i B[2+\alpha_i,\beta_i+\delta_i+1]}\;.
\end{align}
where, $B[a,b]$ denotes the Euler-Beta function. The DEHSS parameterization uses a Variable Flavor Number Scheme (VFNS). In their treatment, the heavy flavor FFs are zero below the quark masses ($m_{c}=1.43\text{ GeV}$ and $m_{b}=4.3\text{ GeV}$) and the splitting functions which mix light and heavy quarks are also set to zero below the heavy quark masses. When the scale is equal to the quark masses, the heavy flavor FFs are parameterized and the splitting functions which mix heavy and light-flavor FFs are introduced. In this paper, we will follow the same treatment of heavy flavor as DEHSS for our nFFs. To parameterize the nFFs, we follow the parameterization in LIKEn \cite{Zurita:2021kli}, where the nuclear modifications to the fragmentation functions are given by
\begin{eqnarray}
\tilde{N}_{i}&\to&\tilde{N}_{i}\Big[1+N_{i,1}(1-A^{N_{i,2}})\Big] \nonumber \\
c_{i}&\to&c_{i}+c_{i,1}(1-A^{c_{i,2}}) \, ,
\label{eq:params}
\end{eqnarray}
where $c\in \left\{\alpha,\beta,\gamma,\delta\right\}$ while 
\begin{align}
    \tilde{N}_i = \frac{N_i}
{B[2+\alpha_i,\beta_i+1]+\gamma_i B[2+\alpha_i,\beta_i+\delta_i+1]}\,.
\end{align}
In the parameterization in LIKEn, the parameters $c_{i,1}$, $c_{i,2}$, $N_{i,1}$, and $N_{i,2}$ represent fit parameters. In our paper, we follow the same parameterization. However, we choose a different set of parameters than LIKEn. Firstly, we note that the SIDIS data is sensitive to the quark TMD FFs at the tree level while the gluon contributions enter only at one loop. Thus these data are mainly sensitive to the quark nTMD FFs. As a result, for both fits that we present, we set the gluon nFFs to be the same as the vacuum gluon FFs. For fit(a) and fit(b), we will see that we cover a different region of $z$ values and are therefore sensitive to different parameters in the collinear nFF.

For the chosen data set of fit(a), we are sensitive to a relatively wide range of $z$ values (0.15 - 0.65), thus we are sensitive to data at both large and small $z$. In our parameterization, the description of the data at extreme values of $z$ is controlled by the $\alpha$ and $\beta$ parameters. Additionally, we note that for this fit, all of the SIDIS data now exists at the HERMES scale $Q^2 \sim 2.4$ GeV$^2$. Thus for this fit, we can simply set $\Gamma = 0$. For fit a, we then choose the parameters
\begin{align}
    \bm{p} = \left\{N_{q1}\,, N_{q2}\,, \alpha_{q1}\,, \alpha_{q2}\,, \beta_{q1}\,, \beta_{q2}\,, g_q^A\,, g_h^A\right\}\,.
\end{align}
For the case of fit(b), we cover a fit of $z$ range of 0.38 to 0.65. Thus for this fit, we are mainly sensitive to parameters that control the region of mid $z$. For this purpose in fit(b), we choose to use the parameters $\delta$ and $\gamma$. Additionally, we note that fit(b) covers a wider range of $Q$ values for the SIDIS data, and thus the simultaneous analysis is sensitive to non-perturbative evolution effects in the nTMD FF. In fit(b), we are left with $9$ parameters
\begin{align}
    \bm{p} =  \left\{N_{q1}\,, N_{q2}\,, \gamma_{q1}\,, \gamma_{q2}\,, \delta_{q1}\,, \delta_{q2}\,, g_q^A\,, g_h^A\,, \Gamma\right\}\,,
\end{align}
while all nuclear modification parameters are set to zero such that these parameters are consistent with the vacuum parameters. Studying the expression for the SIDIS cross section, we see that the nFF must be evolved via a time-like DGLAP evolution from the initial scale to $\mu_i$ by solving the equation
\begin{align}
    \frac{\dd}{\dd{\ln{\mu^2}}}D_{h/q}\left(z,\mu\right) = \left[P^t \otimes D\right] \left(z,\mu\right)\,,
\end{align}
where $P^t$ are the time-like splitting functions which are the same as the space-like ones at LO but differ at NLO. To evolve our parameterization from the initial scale to the scale $\mu_{b_*}$, we use the highly optimized code QCDNUM \cite{Botje:2010ay}, which allows us to treat the heavy flavor contributions to the evolution in the same way as was done in DEHSS. 

\subsection{Numerical recipe for the DIS data}
Each Jefferson Lab data point provides the bin that was used in the variables $x$, $z$, $Q$, and $P_{h\perp}$ to select events. Generating a theoretical prediction for this experimental data then requires integration in each bin. For instance, the numerator of the multiplicities would require the numerical integration
\begin{align}
    d\sigma^A = \int_{x_i}^{x_f} dx \int_{Q^2_i}^{Q^2_f} dQ^2 \int_{z_i}^{z_f} dz \int_{P_{h\perp i}}^{P_{h\perp f}} dP_{h\perp} \frac{d\sigma^A}{dx\, dQ^2\, dz\, dP_{h\perp}}\,.
\end{align}
This multi-dimensional bin integration requires the computation of the cross section at many points and thus massively increases the computation time of the fit. To alleviate this issue in computation time, we take several steps to approximate the bin integration. First, we were supplied the values of $\left\langle x \right \rangle$ and $\left\langle Q \right \rangle$ which are obtained by weighting each event of the Jefferson Lab data as\footnote{We thank Miguel Arratia for providing the bin averaged values}
\begin{align}
    \left \langle x \right \rangle = \int_{x_i}^{x_f} dx x\, \frac{d\sigma^A}{dx\, dQ^2\, dz\, dP_{h\perp}}\,,
    \qquad
    \left \langle Q^2 \right \rangle = \int_{Q^2_i}^{Q^2_f} dQ^2 Q^2\, \frac{d\sigma^A}{dx\, dQ^2\, dz\, dP_{h\perp}}\,.
\end{align}
Second, we approximate the bin integration
\begin{align}
    \int_{z_i}^{z_f} dz \frac{d\sigma^A}{dx\, dQ^2\, dz\, dP_{h\perp}} & \approx \frac{d\sigma^A}{dx\, dQ^2\, d\bar{z}\, dP_{h\perp}} \\
    \int_{P_{h\perp i}}^{P_{h\perp f}} dP_{h\perp} \frac{d\sigma^A}{dx\, dQ^2\, dz\, dP_{h\perp}} & \approx \frac{d\sigma^A}{dx\, dQ^2\, dz\, d\bar{P}_{h\perp}}\,,
\end{align}
where the bar denotes the arithmetic mean of the endpoints of the bin integration. This approximation holds under the assumption that the cross section is slowly varying in the region of integration. In principle the values of $\left \langle z \right \rangle$ and $\left \langle P_{h\perp} \right \rangle$ can also be obtained in the same way that $\left \langle x \right \rangle$ and $\left \langle Q^2 \right \rangle$ to more accurately describe this integration but this would require an additional weighting analysis of the Jefferson Lab events. 

To compute the NLO DIS cross section for the denominator of the multiplicity ratio, we used the APFEL software library \cite{Bertone:2013vaa}. The deuteron cross section was computed using the CT18ANLO parameterization for the PDF using the \textit{SetTargetDIS(``isoscalar")} command while the nuclear cross section was computed using the EPPS21 parameterization. For the case of the Jefferson Lab data, we used the values of $\left\langle x \right \rangle$ and $\left\langle Q \right \rangle$ to obtain the cross section. 

\subsection{Numerical recipe for the Drell-Yan data}
The experimental measurements at the LHC in \cite{ATLAS:2015mwq} and \cite{CMS:2015zlj} require several careful considerations. In these experiments, the transverse momentum distribution of final-state leptons was measured using a 4 TeV proton beam 1.58 TeV per nucleon lead beam. The theoretical formalism presented in the formalism section, as well as our code, are generated in the hadronic CM frame. Due to this asymmetry careful treatment of the LHC data is required. In these measurements, the lab frame and the CM frame are related to one another by a boost of $0.465$ in the direction of the incoming proton. Thus in describing these experimental data, we have offset the values of the rapidity of the final-state lepton pairs to be consistent with the experimental data. In addition to this consideration, we note that the LHC measurements also introduced the fiducial cuts on the final-state leptons. To calculate these fiducial cuts, we use the artemide library \cite{Scimemi:2017etj}. Additionally, to increase the accuracy in describing each data set, we perform bin integration in both $y$ and $Q$ using the interference of the $Z/\gamma^*$. Lastly, we note that the LHC data sets contained an overall luminosity uncertainty. We will discuss the treatment of this uncertainty in the next section.

The Fermilab experiments E866 and E772 measure the transverse momentum distribution of the ratio $R_{AB}$, the Drell-Yan cross section in a heavier nucleus $A$ over that of a lighter baseline nucleus $B$. These experiments were conducted using an 800 $\rm{GeV}$ proton beam.  The E886 measurement uses the baseline nucleus of $B = \rm{Be}$, while the E772 measurement uses a baseline of $B = \rm{D}$. These ratios are measured against $q_{ \perp}$, the transverse momentum of the virtual photon involved in the process. To generate the prediction, we perform a bin integration over $ 0.05 < x_F < 0.3 $ and $5 < Q < 9 $ GeV for the E772 data. For the E866 data, we found it is sufficient to only integrate over $ 0.13 < x_F < 0.93 $ while choosing the arithmetic mean for $Q$ in each reported bin. Lastly, the RHIC PHENIX experiment also measures the transverse momentum distribution of $R_{AB}$ with a baseline of $B = D$ at a center of mass energy $\sqrt{s} = 200$ GeV. We performed a bin integration of $-1.2 < y < -2.2 $, while choosing choosing the arithmetic mean for $Q$ in each reported bin.  \\

\subsection{Fitting procedure}
To perform the fitting procedure for fits a and b, we follow the treatment of \cite{Eskola:2016oht} to minimize the value of
\begin{align}
    \chi^2\left(\left\{\bm{p}\right\}\right) = \sum_i \frac{\left(T_i\left(\left\{\bm{p}\right\}\right)-E_i\right)^2}{\sigma_i^2} + \sum_{j} \left(\frac{1-\mathcal{N}_j\left(\left\{\bm{p}\right\}\right)}{\sigma_{\rm norm}}\right)^2\,.
\end{align}
In the first term on the right-hand side of this expression, we sum over all data points, which are indexed using $i$. In this term $T_i$ and $E_i$ are used to denote the value of the theory at some particular parameter values $\bm{p}$ and the kinematics of point $i$, and experimentally measured value of the point $i$. Additionally, we denote the quadrature sum of the statistical and systematic errors as $\sigma_i$. In the second term of this expression, we include a sum over $j$, where j runs over the CMS and ATLAS data sets. There $\mathcal{N}_j$ is used to denote the introduction of the normalization while $\sigma_{\rm norm}$ is used to denote the luminosity uncertainty of those data sets. This second term serves to allow us to change the overall normalization of the CMS and ATLAS data sets but penalizes the value of the $\chi^2$ when normalization differs from $1$. 

At the beginning of the fit, we use a random number generator to assign an initial value to each parameter in the fit. However, we note that for certain parameter values, the FFs become non-integrable, which would call into question the probability density interpretation of the nFF. To avoid this issue, we defined end-points for each parameter such that for any value initialized within that range and for all $A$ in the fit, the nFFs are integrable. The $\chi^2$ is then minimized using a Migrad minimization procedure through the Minuit software library \cite{James:1994vla}. 

\subsection{Treatment of uncertainties}
We consider two sources of uncertainty in our analyses. First, we need to characterize the uncertainty that is associated with the collinear nFF and the parameters which characterize the nuclear modification to the TMDs. For simplicity, we will refer to the total uncertainty associated with these parameters as the `fit uncertainty'. To generate the fit uncertainties, we use the \textit{replica} method \cite{Bacchetta:2017gcc}. In this method, the central value and uncertainty of each data point are recorded into a one-dimensional array
\begin{align}
    \bm{E}_0 = \left\{E_1, E_2,  .... E_N\right\}\,,
    \qquad
    \bm{\sigma}_0 = \left\{\sigma_1, \sigma_2,  .... \sigma_N\right\}\,,
\end{align}
where $N$ is the number of data in the fit(a) and the $0$ is used to denote that this is the original data set. For each data point $i$, the replica method then generates a random number from a Gaussian distribution of width $\sigma_i$. To clarify, we will denote this random number $r(\sigma_i)$. Thus for some replica $\alpha$, we have the central values and experimental uncertainties
\begin{align}
    \bm{E}_\alpha = \left\{E_1+r\left(\sigma_1\right), E_2+r\left(\sigma_2\right), .... E_N+r\left(\sigma_N\right)\right\}\,,
    \qquad
    \bm{\sigma}_\alpha = \left\{\sigma_1, \sigma_2, .... \sigma_N\right\}\,.
\end{align}
In our analysis, we generate 200 replicated sets of experimental data. For each set of replicas, we perform a $\chi^2$ minimization using the procedure that was outlined in the previous section. After performing this procedure, we will arrive at a set of parameters for each replica $\bm{p}_\alpha$. To characterize the fit uncertainty in all of the below plots, we begin by generating a prediction for each set of parameters to obtain the vector
\begin{align}
    \bm{\mathcal{O}} = \left\{\mathcal{O}\left(\bm{p}_1\right), \mathcal{O}\left(\bm{p}_2\right), ... \mathcal{O}\left(\bm{p}_{200}\right)\right\}\,,
\end{align}
where $\mathcal{O}\left(\bm{p}_{\alpha}\right)$ is some prediction which depends on the parameter set $\bm{p}_\alpha$. The central value and fit uncertainties are then obtained by calculating the average and standard deviation of $\bm{\mathcal{O}}$. 

In addition to the fit uncertainties, we must also identify the uncertainties of the PDF and nPDF. To address this, we note that EPPS21 provides 48 error sets which characterize the uncertainty in their analysis as well as 58 error sets characterize the size of the proton PDFs uncertainties stemming from CT18ANLO. Following the EPPS21 prescription, the $90\%$ confidence interval above and below the central curve can be generated using
\begin{align}
    \left(\Delta X^{+}\right)^2 \approx \sum_{k=1}^{53}\left[\max \left\{X\left(S_k^{+}\right)-X\left(S^0\right), X\left(S_k^{-}\right)-X\left(S^0\right), 0\right\}\right]^2\,,
    \\
    \left(\Delta X^{-}\right)^2 \approx \sum_{k=1}^{53}\left[\min \left\{X\left(S_k^{+}\right)-X\left(S^0\right), X\left(S_k^{-}\right)-X\left(S^0\right), 0\right\}\right]^2\,,
\end{align}
where the uncertainty above and below the curve is given by $\Delta X^{\pm}$. 
In these expressions, $S^0$ represents the prediction of the central set, and $S^{\pm}_k$ represents the 2 error sets in the k direction. Full collinear uncertainty is just the sum of collinear errors in each error direction k. To get the collinear uncertainty at 68\% confidence level, we divide the collinear uncertainty at 90\% confidence level by 1.645. To generate the total fit uncertainty, we now add the fit and EPPS21 uncertainties in quadrature.

\section{Results}
\subsection{Fit(a)}\label{subsec:Fita}
\begin{table}[H]
\centering
\begin{tabular}{c c c c c c}
\hline
\hline
Collaboration  & Process  & Baseline  & Nuclei  & $\mathrm{N}_\mathrm{data}$  & $\chi^2$ \\
\hline
HERMES~\cite{Airapetian:2007vu} & SIDIS($\pi$)  & D  & Ne, Kr, Xe & 47 & 8.8\\
RHIC~\cite{Leung:2018tql}  & DY & p & Au & 4 & 1.5 \\
E772~\cite{Alde:1990im}  & DY  & D  & C, Fe, W & 16 & 21.0\\
E866~\cite{Vasilev:1999fa}  & DY & Be & Fe, W & 28 & 35.1\\
CMS~\cite{CMS:2015zlj}  & $\gamma^*/Z$ & N/A & Pb & 8 & 10.3\\
ATLAS~\cite{ATLAS:2015mwq}  & $\gamma^*/Z$ & N/A & Pb & 7 & 13.2\\
Total & & & & 110 & 89.7\\
\hline
\hline
\end{tabular}
\caption{The $\chi^2$ for fit(a). The values listed are the central fit.}
\label{tab:chi2-a}
\end{table}

In Tab.~\ref{tab:chi2-a}, we provide the $\chi^2$ for each data set using the parameter values of the central fit. In this table, we include the process, the light nucleus which is used as a baseline, the heavy nucleus, the number of data, and the $\chi^2$ for each data set. Using the fitting procedure outlined in the previous section, we obtain a $\chi^2/\rm{d.o.f}$ of $0.879$.

In Tab.~\ref{tab:Params-fita}, we include the values of the parameters obtained in fit(a). The central value for each parameter was obtained by averaging the parameter value for each replica while the uncertainty is obtained by measuring the mean positive and negative distances. From top to bottom, the table contains the obtained $q1$ parameter values, the $q2$ parameter values, and the parameter values which characterize the nuclear modification to the transverse momentum. From this table, we see that the parameter $\beta_{q2}$ is consistent with zero, which suggests that additional experimental data is required to constrain the medium modification to the FF. Additionally, we see that the current set of experimental data suggests that the value of $g_q^A$ is close to $g_h^A$ and that both values are consistent with our previous analysis in \cite{Alrashed:2021csd}.

\begin{table}[H]
\centering
\begin{tabular}{c c c}
\hline
\hline
$N_{q1} = 0.238^{+0.365}_{-0.0789}$ & $\alpha_{q1} = 0.082^{+1.12}_{-0.0393}$ & $\beta_{q1} = 0.00169^{+0.176}_{-0.283}$ \\
$N_{q2} = 0.259^{+0.0772}_{-0.103}$ & $\alpha_{q2} = 0.388^{+0.0799}_{-0.270}$ & $\beta_{q2} = 1.030^{+0.512}_{-1.11}$ \\
$g_q^A = 0.016 ^{+0.00187}_{-0.00189}$ & $g_h^A = 0.013 ^{+0.0104}_{-0.00730}$ & \\
\hline
\hline
\end{tabular}
\caption{Parameter values for fit(a).}
\label{tab:Params-fita}
\end{table}

\begin{figure}[H]
    \centering
    \includegraphics[valign = c, width = 0.44\textwidth]{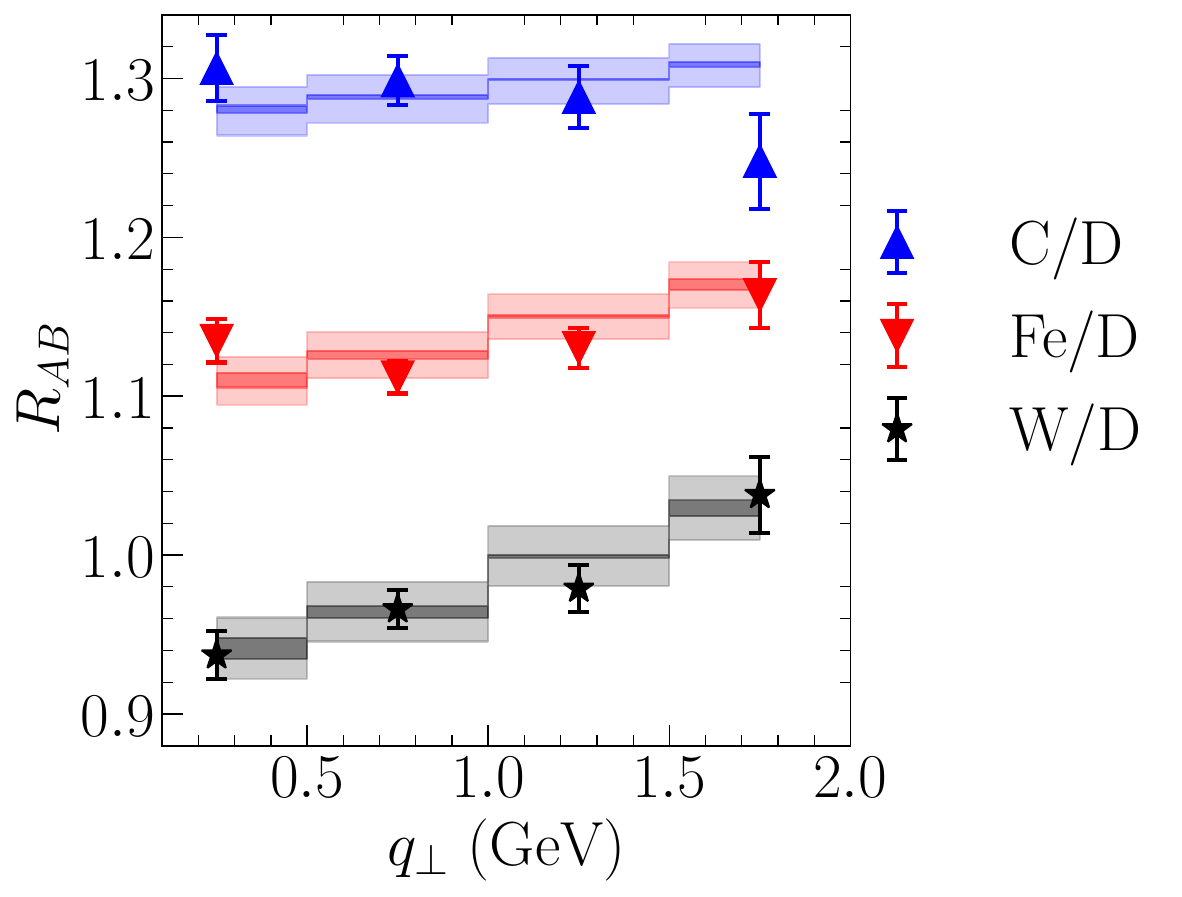} 
    \includegraphics[valign = c, width=0.44\textwidth,]{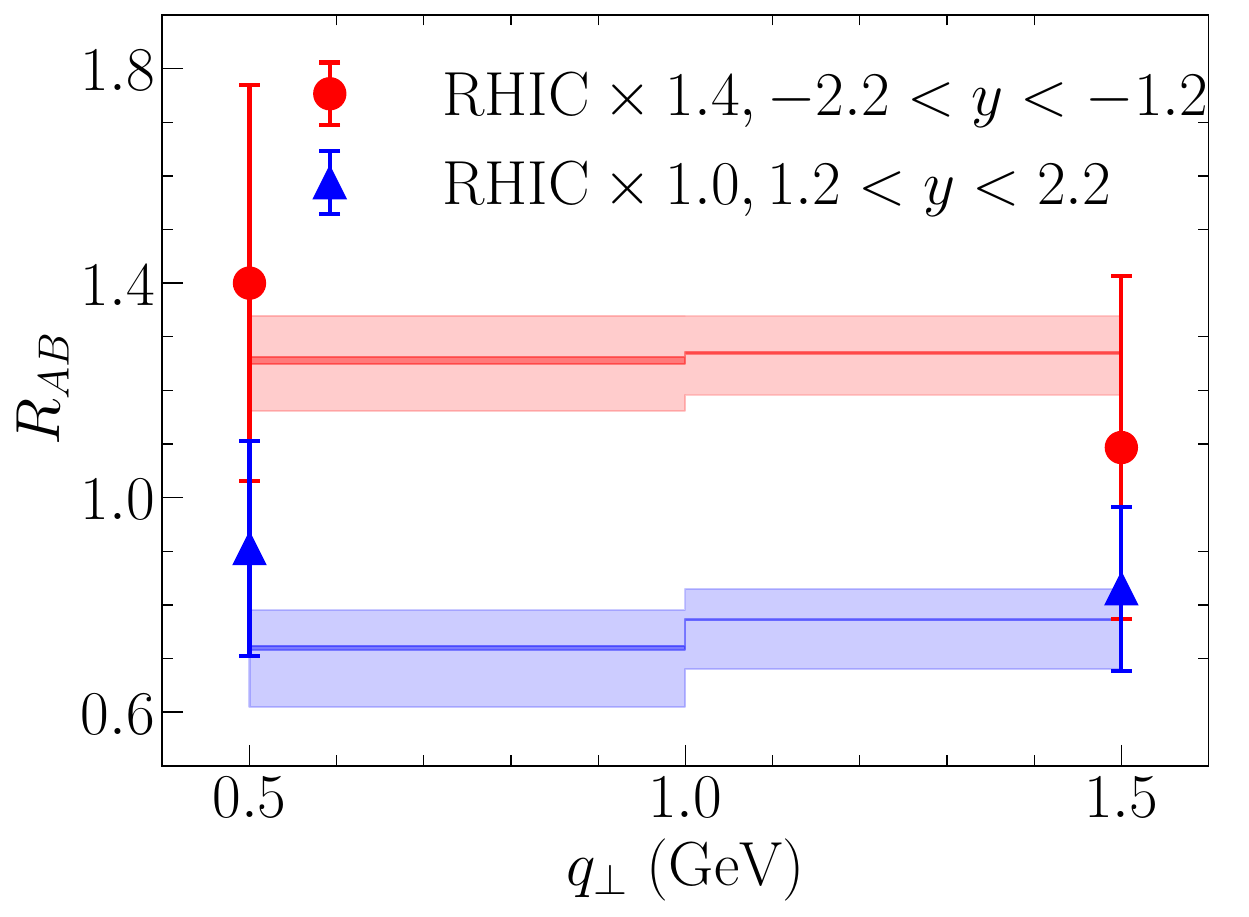} 
    \caption{Left: The description of the E772 data. Right: The description of the RHIC data. The E772, C/D and Fe/D data have been multiplied by factors of 1.3 and 1.15 respectively.}
    \label{fig:E772-RHIC-a}
\end{figure}
\begin{figure}[H]
    \centering
    \includegraphics[width = 1.0\textwidth]{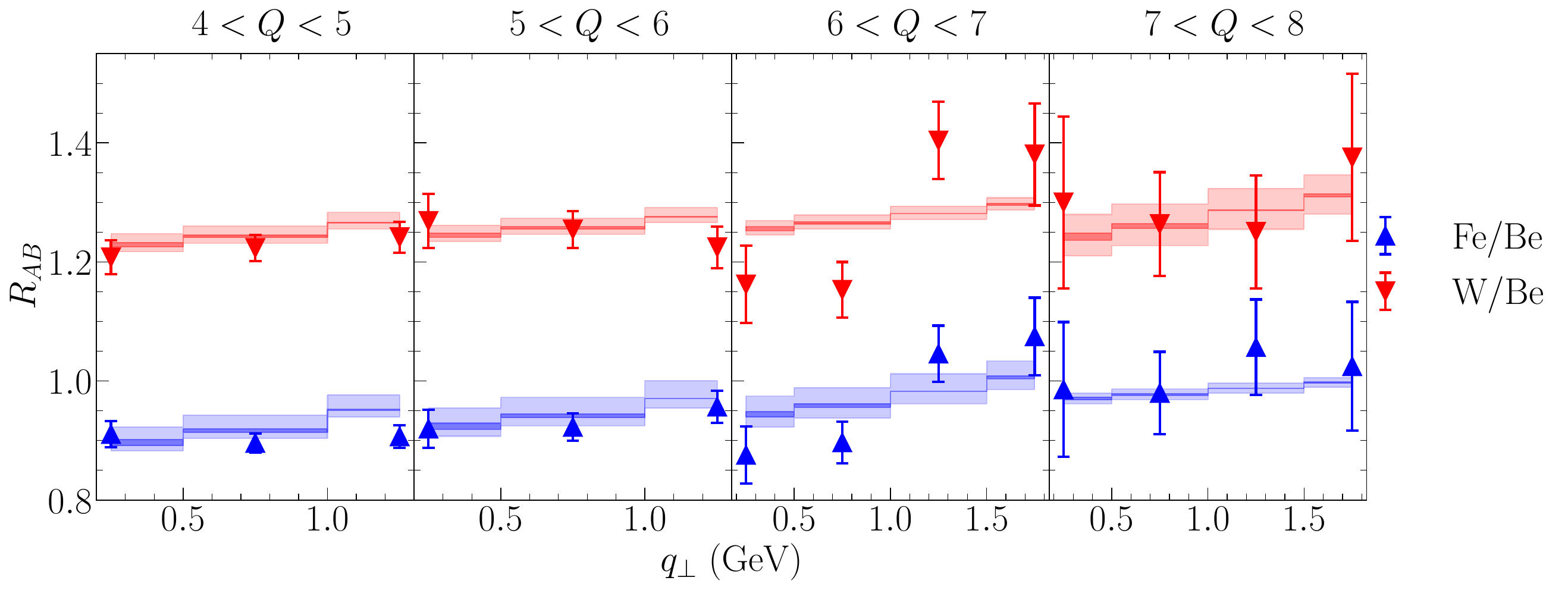} 
    \caption{Description of the E866 data set. The Fe/Be data has been multiplied by a factor of 1.3.}
    \label{fig:E866-a}
\end{figure}
\begin{figure}[H]
    \centering
    \includegraphics[width = 0.4\textwidth]{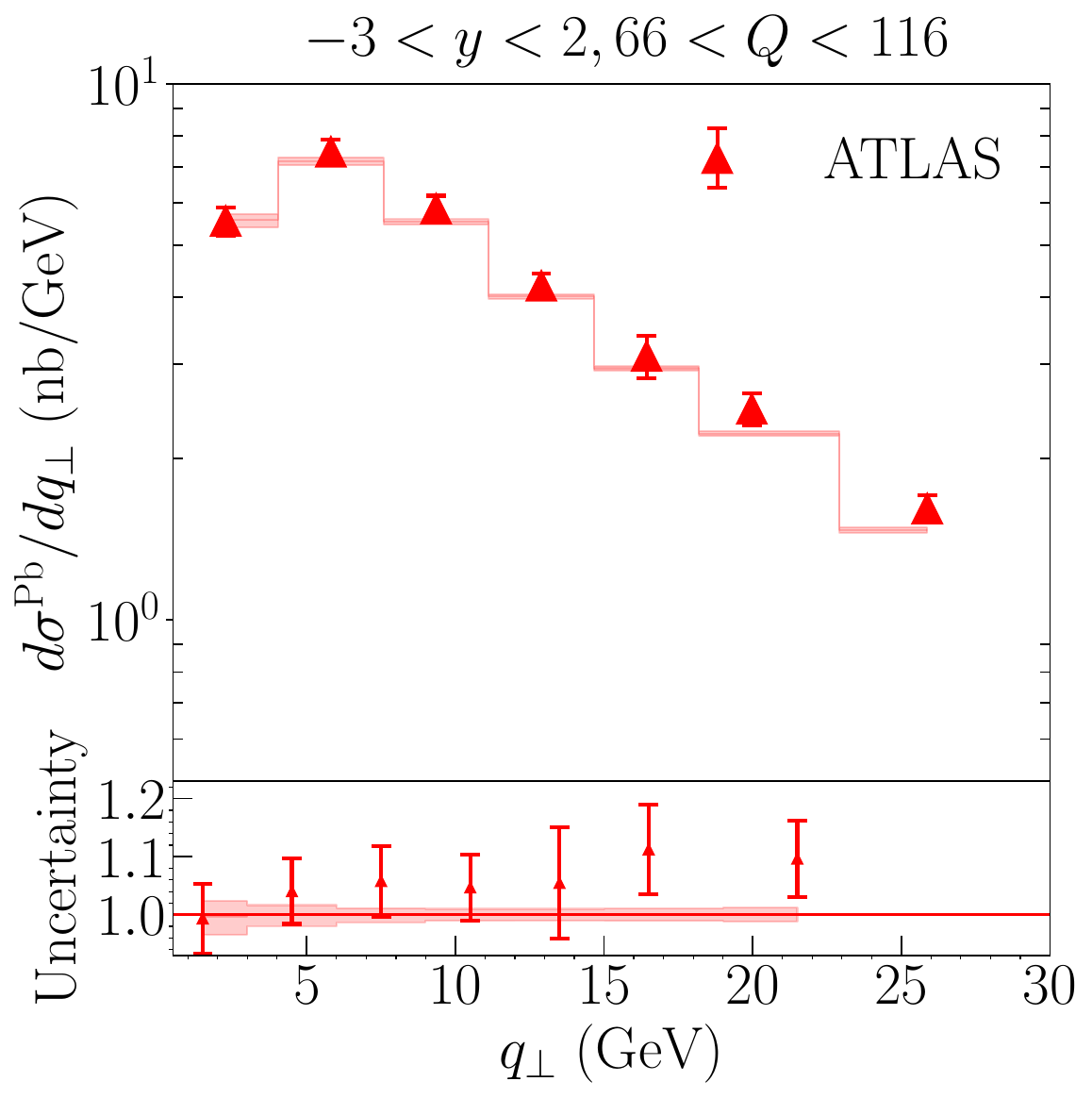} 
    \includegraphics[width = 0.4\textwidth]{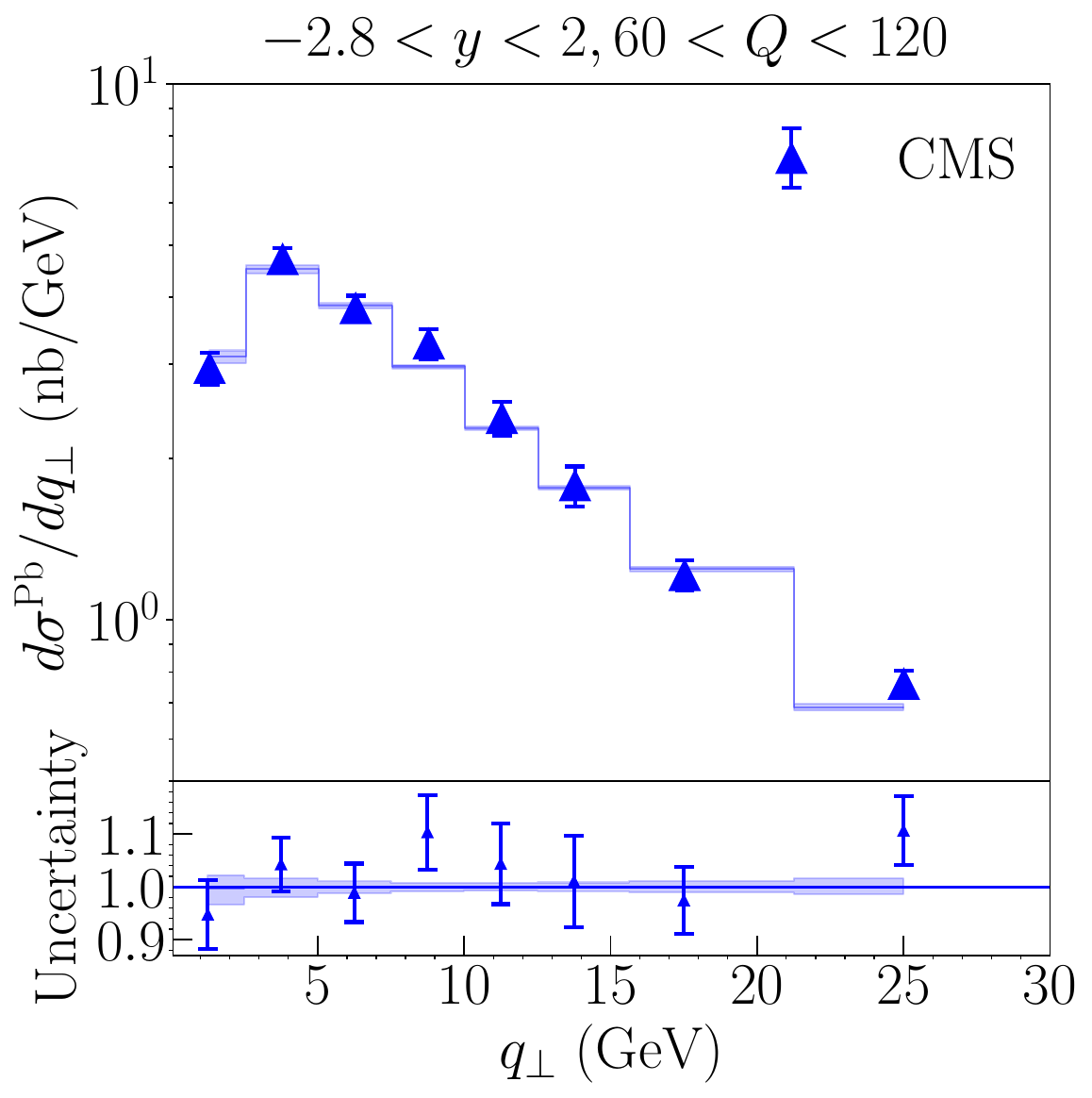} 
    \caption{Left: The description of the ATLAS dataset. Right: The description of the CMS dataset. In the bottom plots, we divide the theory and the uncertainty band by the central theory curve.}
    \label{fig:LHC-a}
\end{figure}
\begin{figure}[H]
    \centering
    \includegraphics[width = 1\textwidth]{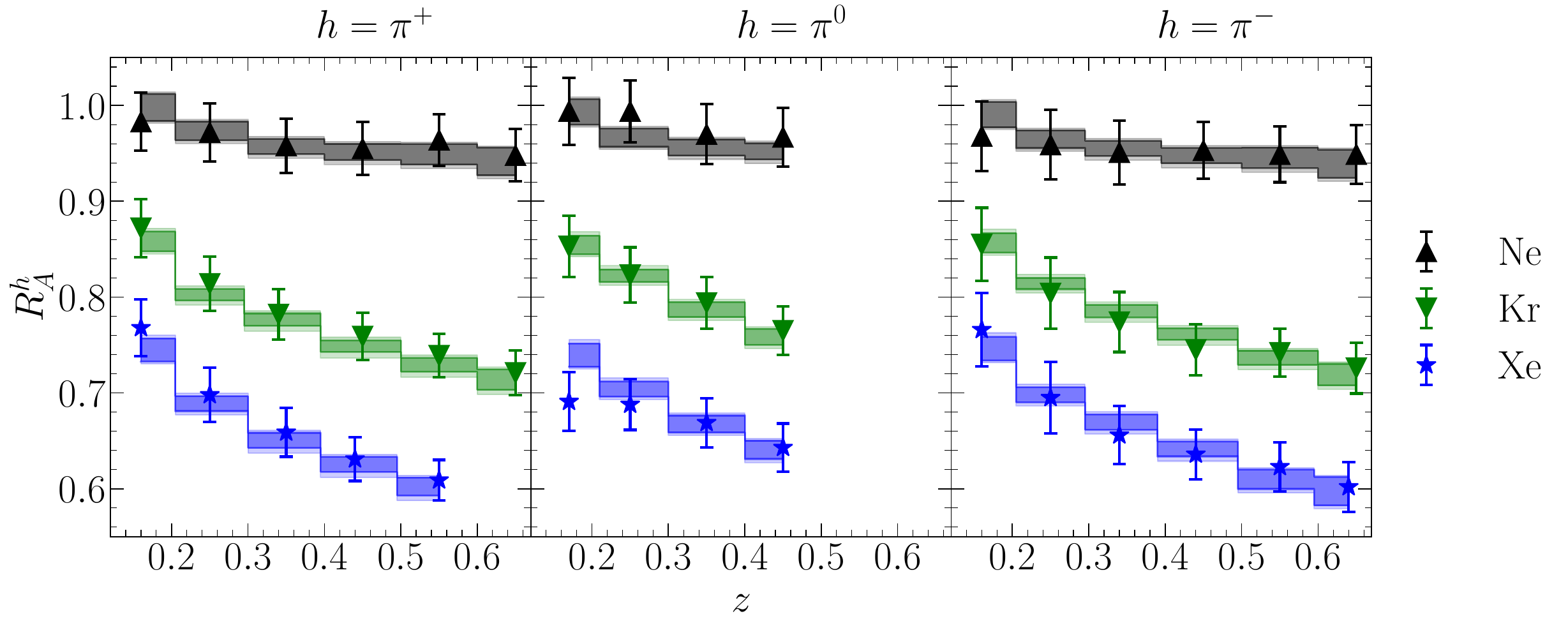}
    \caption{The description of HERMES dataset. The Xe data has been offset downward by 0.07, while the Ne data has been offset upward by 0.07.}
    \label{fig:HERMES-a}
\end{figure}

In Figs.~\ref{fig:E772-RHIC-a}, \ref{fig:E866-a}, \ref{fig:LHC-a}, and \ref{fig:HERMES-a}, we plot the description of the experimental data. The dark bands represent the fit uncertainties while the light band represents the uncertainty from the nPDF and the PDF. As the description of the Drell-Yan data depends only on a single fit parameter, we can see in Figs.~\ref{fig:E772-RHIC-a}, \ref{fig:E866-a}, and \ref{fig:LHC-a} that the nPDF uncertainties are much larger than the fit uncertainty. However, the description of the HERMES data is controlled by 7 parameters and thus the uncertainties in Fig.~\ref{fig:HERMES-a} are dominated by the fit uncertainties. 

From the E772 and E866 cross section ratios in Figs.~\ref{fig:E772-RHIC-a} and \ref{fig:E866-a}, we see the effects of the broadening in the data and the parameterizations, which are controlled by the $g_q^A$ parameter. The definite broadening of the experimental data sets is responsible for the $g_q^A$ parameter being inconsistent with zero within the parameter uncertainties. In Fig.~\ref{fig:LHC-a}, we see that the size of the fit uncertainties is much smaller than the experimental uncertainties. This is due to the small uncertainties in the nPDF in the $x$ region that is spanned by the LHC data as well as the transverse momentum of the incoming partons being dominated by perturbative radiation. 

\begin{figure}[H]
    \centering
    \includegraphics[width = 0.49\textwidth]{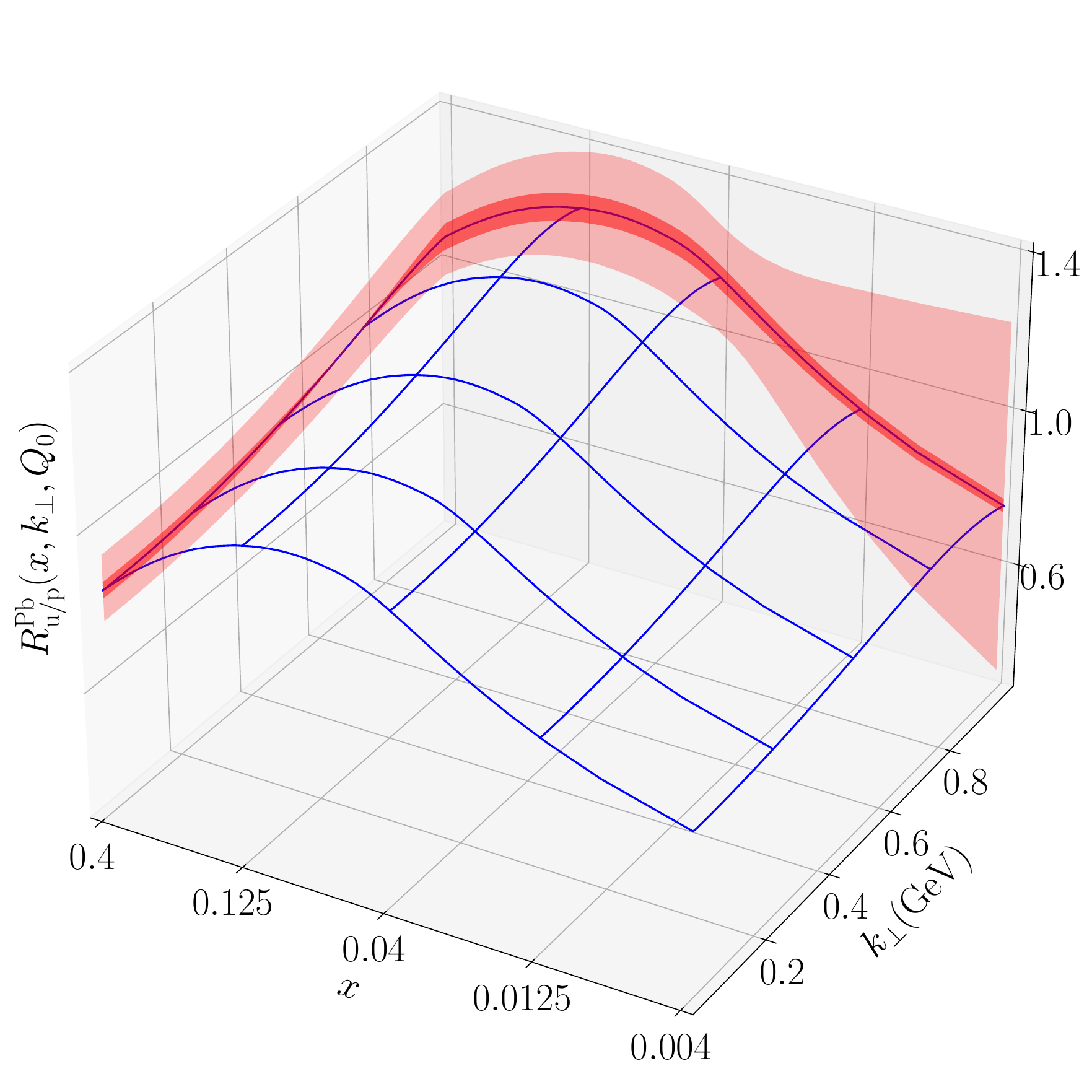}
    \includegraphics[width = 0.49\textwidth]{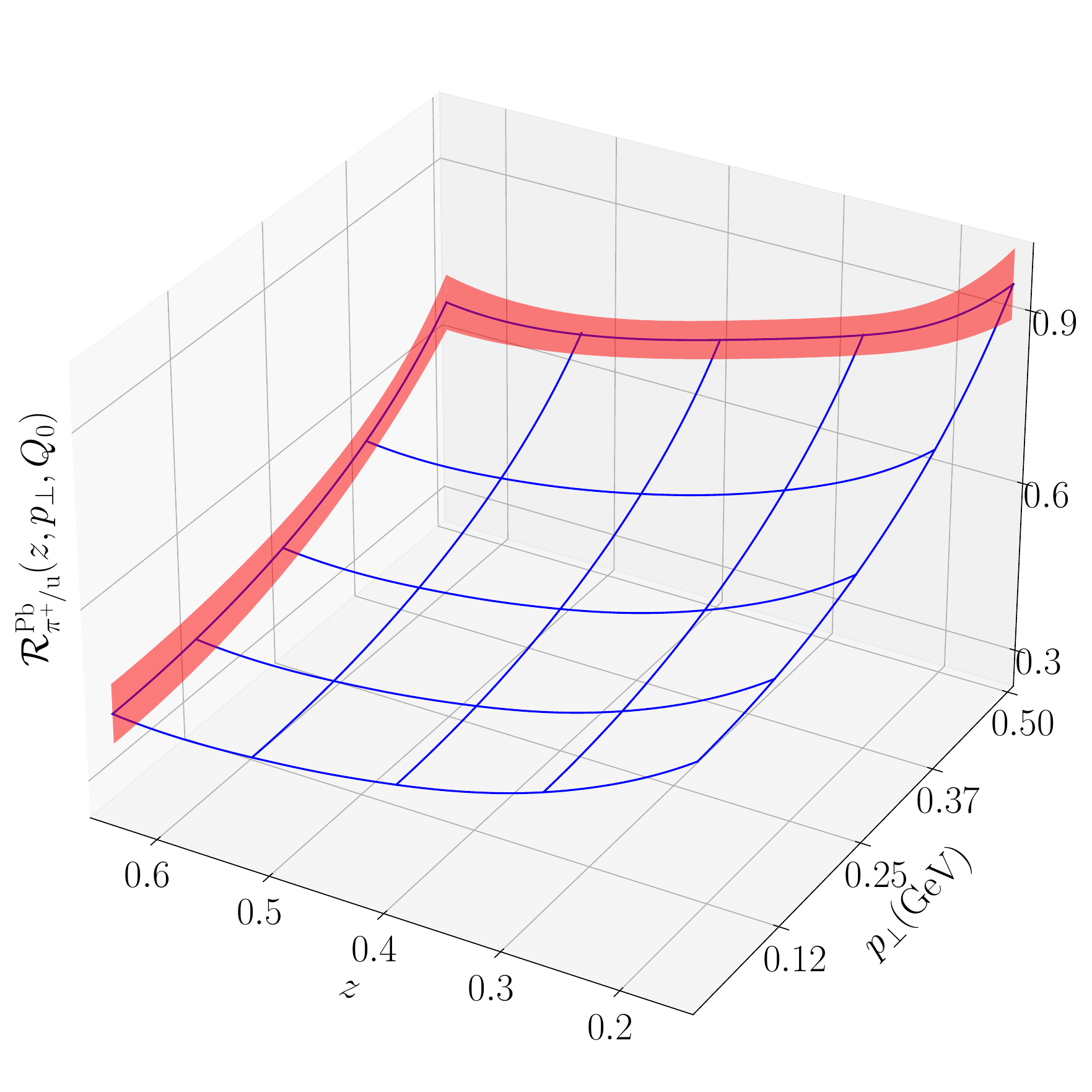}    
    \caption{Left: the extracted nuclear ratio for the  TMD PDF at $Q_0 = \sqrt{2.4}$ GeV. Right: the extracted nuclear ratio for the TMD FF at the same scale. }
    \label{fig:3Dfita-a}
\end{figure}

In the left side of Fig.~\ref{fig:3Dfita-a}, we plot the following ratio
\begin{align}\label{eq:RPb}
    R_{u/p}^{\rm Pb}\left(x,k_\perp,Q_0\right) = \frac{f_{q/p/\rm{Pb}}\left(x,k_\perp,Q_0,Q_0^2\right)}{f_{q/p}\left(x,k_\perp,Q_0,Q_0,Q_0^2\right)}
\end{align}
which gives the ratio of the nTMD PDF in a proton that is bound in a Pb nucleus to that of an unbound proton at $Q_0 = \sqrt{\zeta_0}$. In that plot, we see that for lines of constant $k_\perp$ the function demonstrates the behavior of the nPDF, namely the shadowing, anti-shadowing, and EMC effects. The lines of constant $x$ grow as we increase the transverse momentum of the quarks, which is governed by the broadening parameter $g_q^A$. In this figure, the dark band represents the fit uncertainty, which is controlled only by the $g_q^A$ parameter. In the light band, we plot the nPDF uncertainty. On the right side of this figure, we plot the ratio
\begin{align}\label{eq:RrPb}
    \mathcal{R}_{\pi^+/u}^{\rm Pb}\left(z,p_\perp,Q_0\right) = \frac{D_{\pi^+/u}^{\rm Pb}\left(z,p_\perp,Q_0,Q_0^2\right)}{D_{\pi^+/u}\left(z,p_\perp,Q_0,Q_0^2\right)}
\end{align}
for the TMD FF. We see that at large values of $z$, there is a suppression of the nFF, while at small values of $z$, there is an enhancement in the ratio. This trend is suggested by the HERMES data in Fig.~\ref{fig:HERMES-a} and was also present in the LIKEn extraction. The dark band in this figure represents the fit uncertainty. 

\begin{figure}[H]
    \centering
    \includegraphics[width = \textwidth]{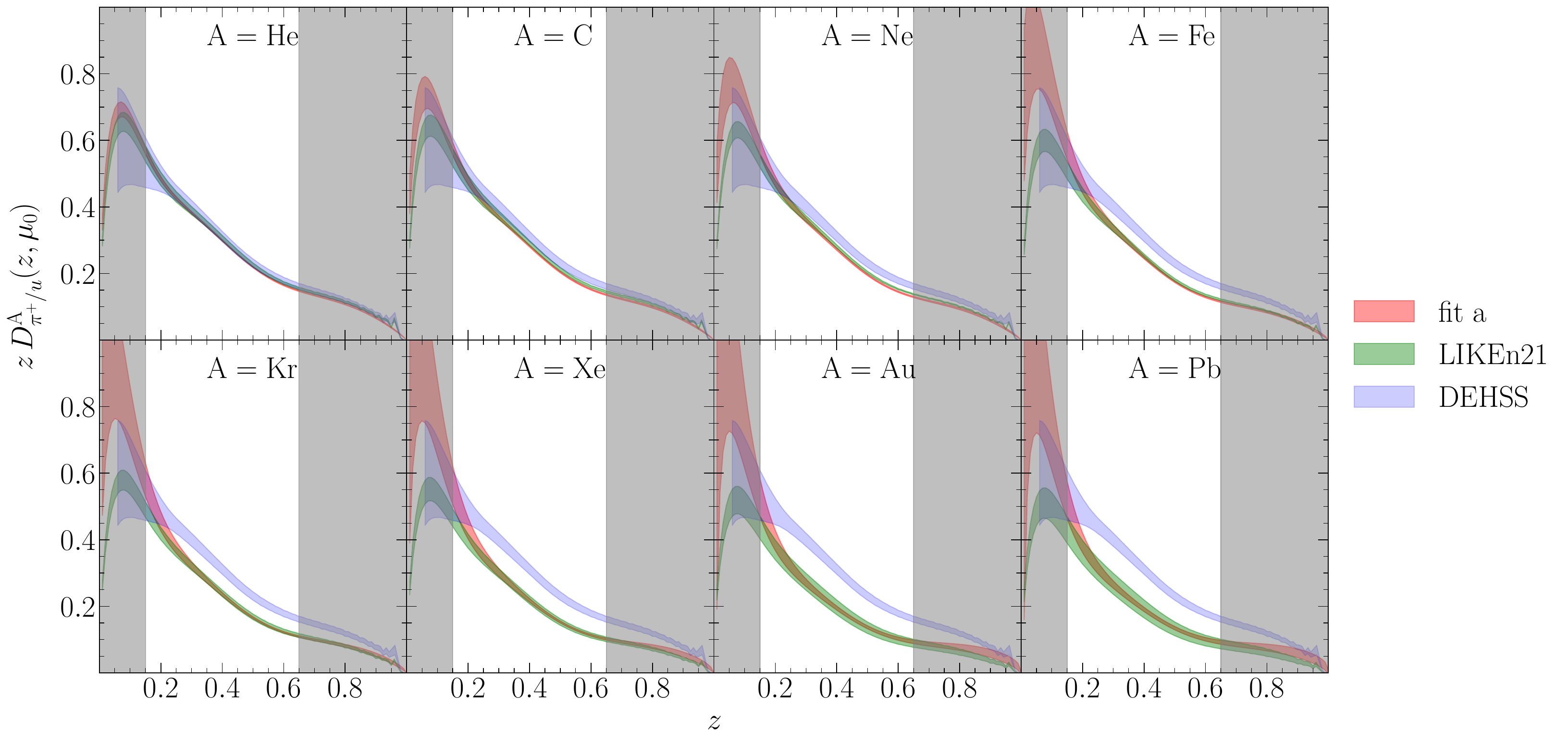}
    \caption{The uncertainties of the FFs at the initial scale $\mu_0 = 1$ GeV. The red, green, and blue bands represent the uncertainties in fit (a), LIKEn, and the vacuum DEHSS FF. The grey region represents the region where we did not have data.}
    \label{fig:Comparison-a}
\end{figure}

In Fig.~\ref{fig:Comparison-a}, compare our extracted nFF against LIKEn and DEHSS. The LIKEn uncertainties in green were generated using the error sets provided for that extraction along with the procedure outlined in the paper. In red, we plot the uncertainty from fit(a) and in blue we plot the uncertainties of DEHSS. The grey region represents the region where fit(a) did not have data. We see that in the region where we had HERMES data that the LIKEn and fit(a) extractions were consistent with one another for all $A$. While we see that in the grey region, the two extractions differ from one another. Additionally, we see that while both LIKEn and our extraction agree with DEHSS for small $A$, they disagree with DEHSS for large nuclei as expected. 

\subsection{Fit(b)}\label{subsec:Fitb}
\begin{table}[H]
\centering
\begin{tabular}{c c c c c c}
\hline
\hline
Collaboration  & Process  & Baseline  & Nuclei  & $\mathrm{N}_\mathrm{data}$  & $\chi^2$ \\
\hline
JLAB~\cite{CLAS:2021jhm} & SIDIS($\pi$)  & D & C, Fe, Pb & 36 & 41.7\\
HERMES~\cite{Airapetian:2007vu} & SIDIS($\pi$)  & D  & Ne, Kr, Xe & 18 & 10.2\\
RHIC~\cite{Leung:2018tql}  & DY & p & Au & 4 & 1.3 \\
E772~\cite{Alde:1990im}  & DY  & D  & C, Fe, W & 16 & 40.2\\
E866~\cite{Vasilev:1999fa}  & DY & Be & Fe, W & 28 & 20.6\\
CMS~\cite{CMS:2015zlj}  & $\gamma^*/Z$ & N/A & Pb & 8 & 10.4\\
ATLAS~\cite{ATLAS:2015mwq}  & $\gamma^*/Z$ & N/A & Pb & 7 & 13.3\\
Total & & & & 117 & 137.8\\
\hline
\hline
\end{tabular}
\caption{The $\chi^2$ for fit(b). The values listed are for the central fit.}
\label{tab:chi2-b}
\end{table}

In Tab.~\ref{tab:chi2-b}, we provide the $\chi^2$ and the description of each data set. We note that the number of HERMES data has changed from Tab.~\ref{tab:chi2-a} and Tab.~\ref{tab:chi2-b} due to the different cuts used in each fit and because we use the $P_{h\perp}$ projection of the data. In total for fit(b), we obtain a $\chi^2/\rm{d.o.f}$ of $1.275$. 

\begin{table}[H]
\centering
\begin{tabular}{c c c}
\hline
\hline
$N_{q1} = 0.256^{+1.07}_{-0.194}$ & $\gamma_{q1} = 0.006^{+0.727}_{-0.873} $ & $\delta_{q1} = 0.184^{+0.883}_{-0.340} $ \\
$N_{q2} = 0.156^{+0.137}_{-0.0906} $ & $\gamma_{q2} = 1.150^{+0.334}_{-0.783} $ & $\delta_{q2} = 0.474^{+0.144}_{-0.232} $ \\
$\Gamma = 2.200^{+0.135}_{-0.0925} $ & $g_q^A = 0.440^{+0.0461}_{-0.0323} $ & $g_h^A = 0.038^{+0.0157}_{-0.0150} $\\
\hline
\hline
\end{tabular}
\caption{The parameter values for fit(b)}
\label{tab:params-b}
\end{table}

The parameter values obtained from fit(b) are given in Tab.~\ref{tab:params-b} where the central value and the uncertainties are obtained by taking the average and the mean positive/negative distances once again. We see from this table that the $\gamma_{q1}$ and $\delta_{q1}$ parameter values are consistent with zero, once again emphasizing the need for additional data. We see that the central value of $g_q^A$ is much larger than the central value given in fit(a). This effect originates from the parameterization that is used for the TMD physics. Namely, since Drell-Yan data sets tend to have large $Q$ values, the broadening is suppressed for these data sets by the factor $(Q_0/Q)^\Gamma$. Thus a larger value of $g_q^A$ is required to describe these data. However as the SIDIS data tends to be at small $Q$, we see that the value of the parameter $g_h^A$ is of the same order as the value obtained from fit(a). 

\begin{figure}[H]
    \centering
    \includegraphics[valign = c, width = 0.44\textwidth]{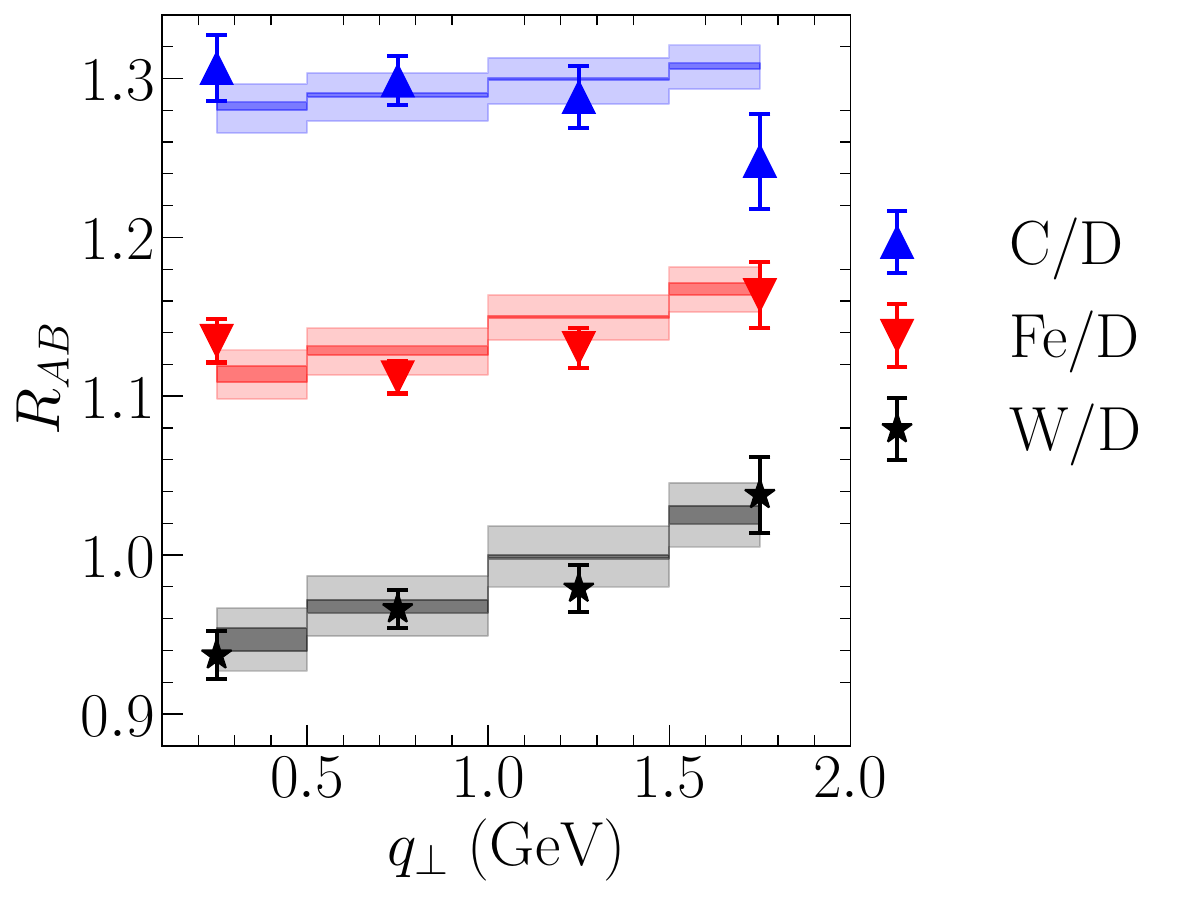} 
    \includegraphics[valign = c, width=0.44\textwidth,]{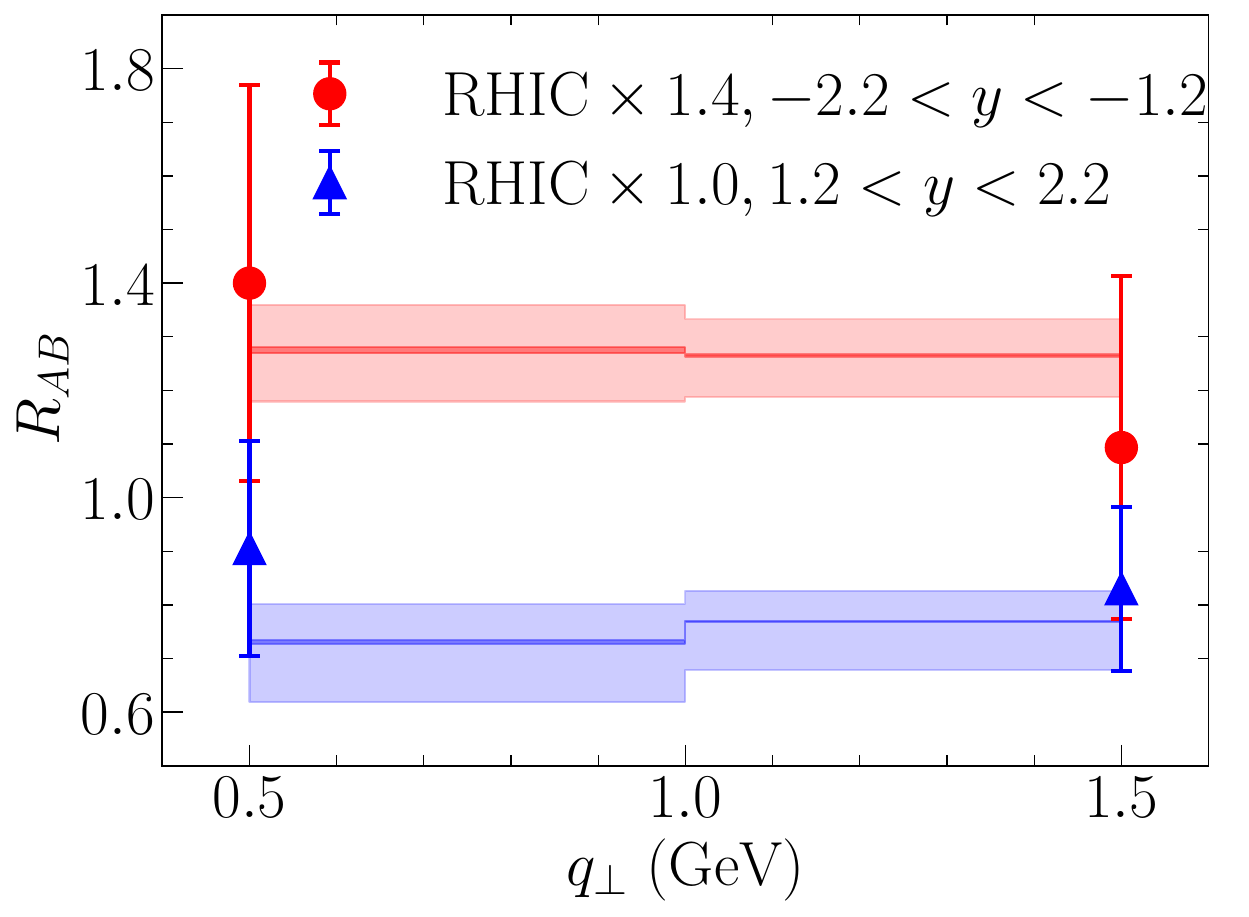} 
    \caption{Left: The description of the E772 data. Right: The description of the RHIC data. For E772, the C/D and Fe/D data have been multiplied by factors of 1.3 and 1.15 respectively.}
    \label{fig:E772-RHIC-b}
\end{figure}

\begin{figure}[H]
    \centering
    \includegraphics[width = 1.0\textwidth]{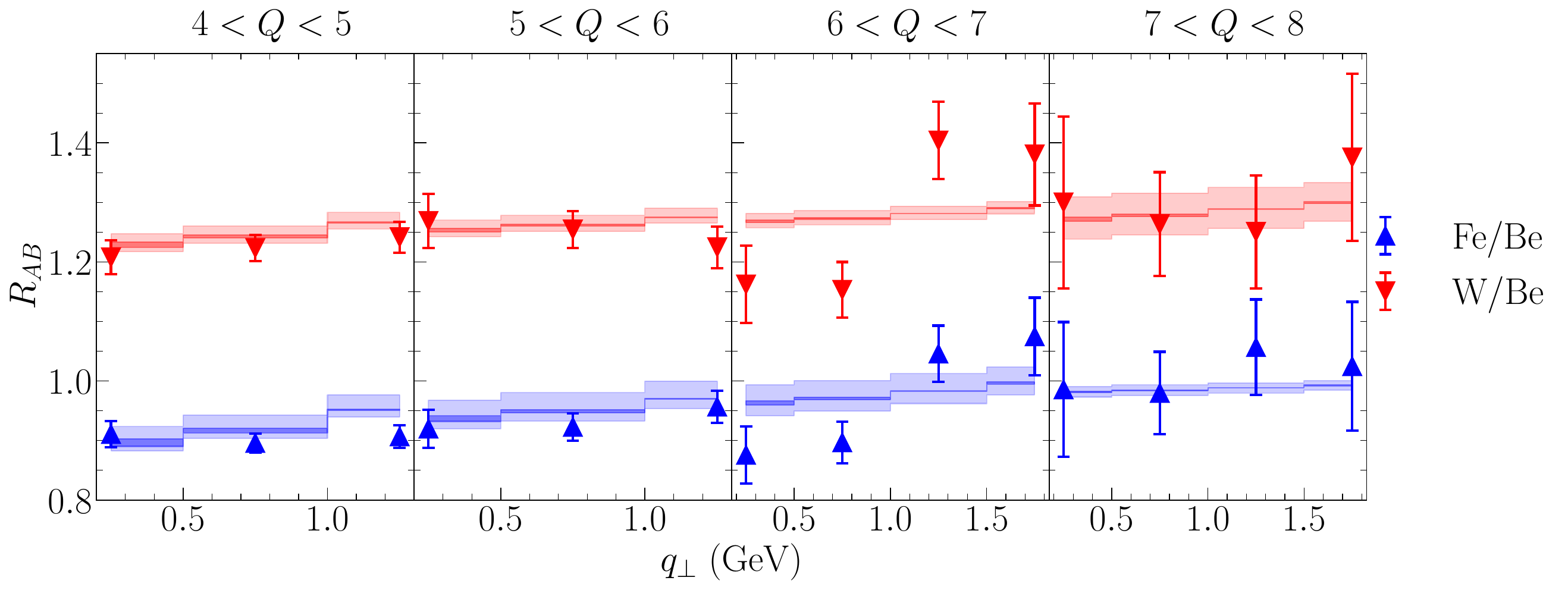} 
    \caption{Description of the E866 data set. The Fe/Be data has been multiplied by a factor of 1.3.}
    \label{fig:E866-b}
\end{figure}

\begin{figure}[H]
    \centering
    \includegraphics[width = 0.4\textwidth]{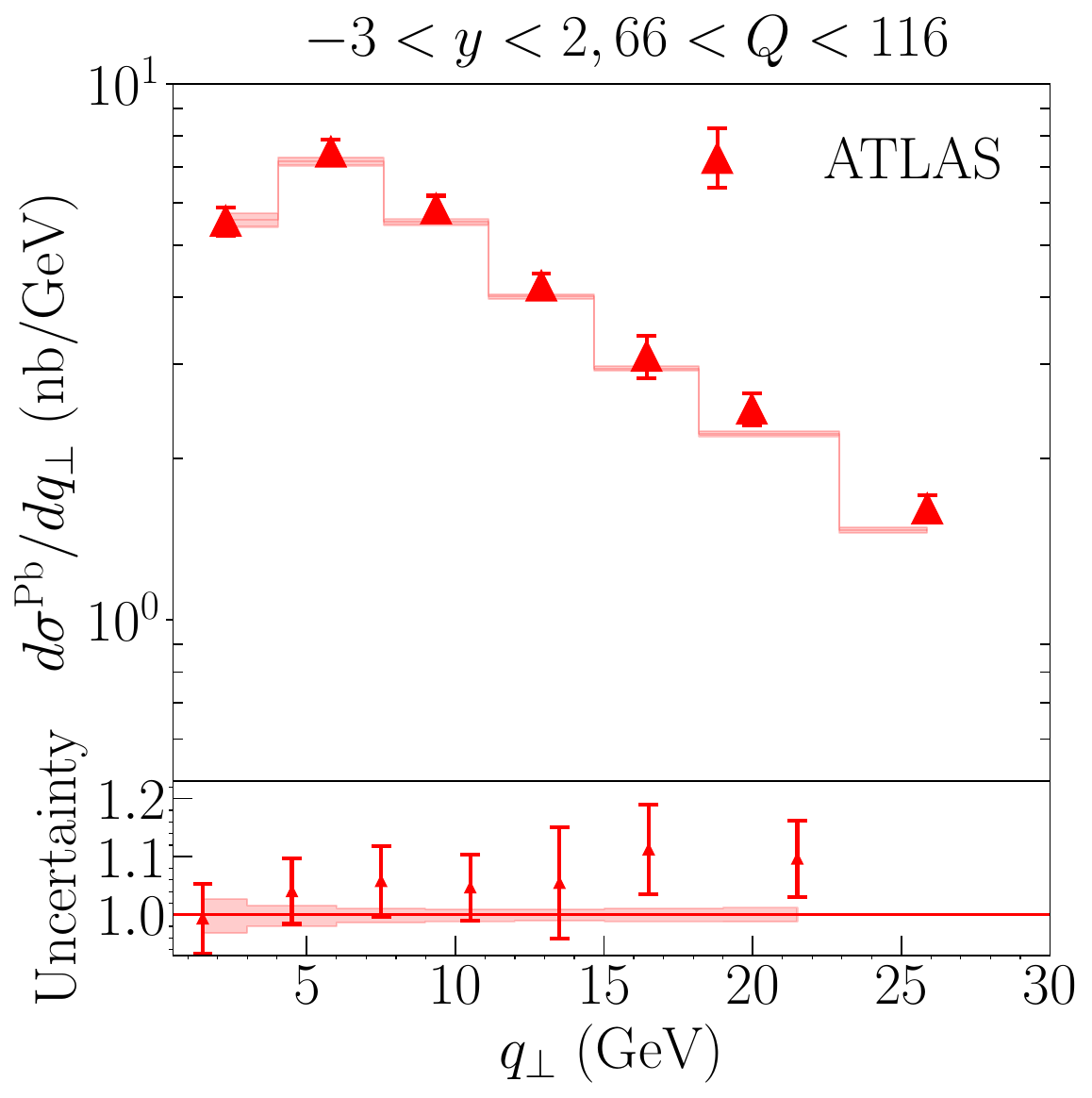} 
    \includegraphics[width = 0.4\textwidth]{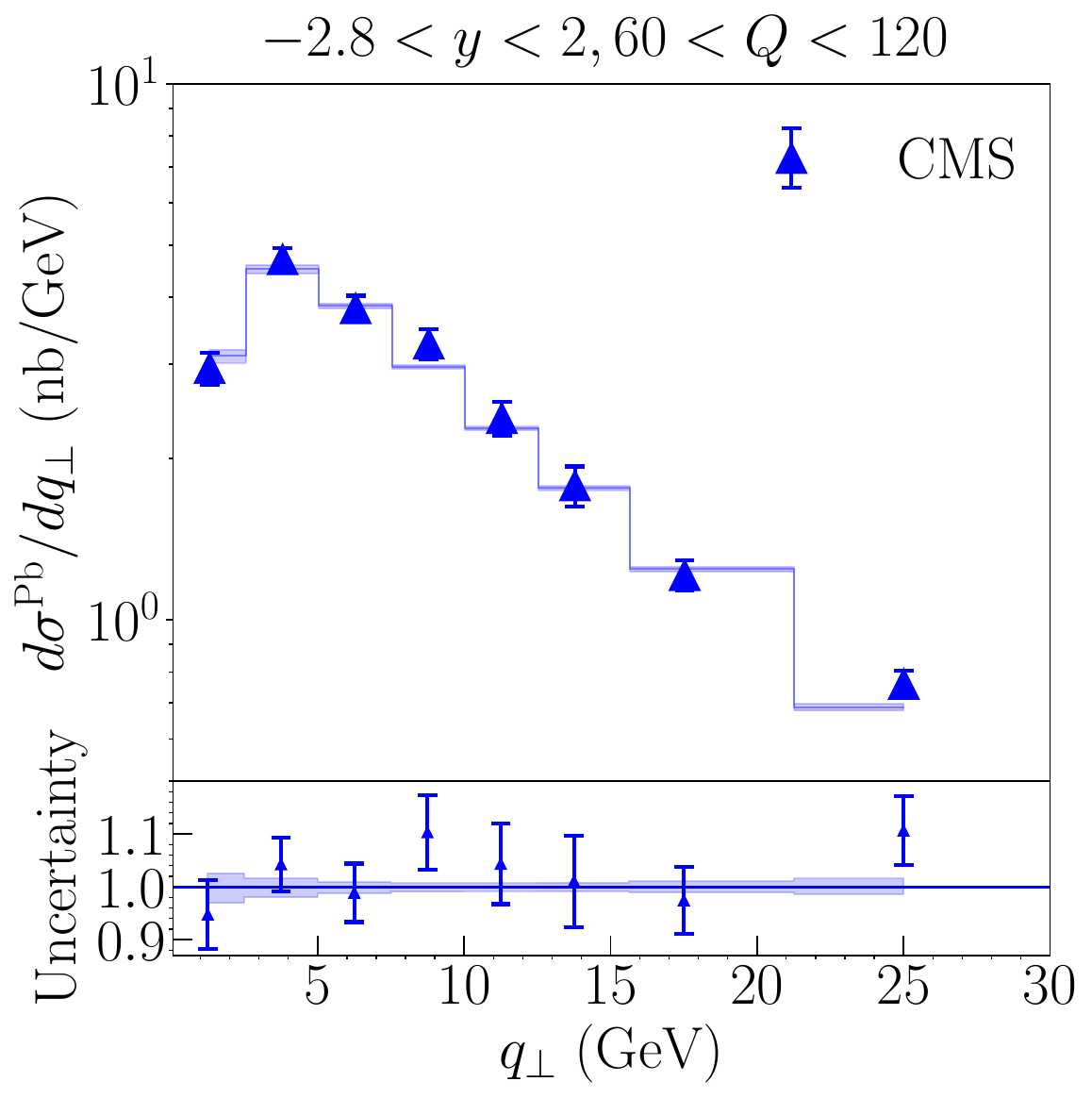} 
    \caption{Left: The description of ATLAS dataset. Right: The description of the CMS dataset. In the bottom plots, we divide the theory and the uncertainty band by the central theory curve.} 
    \label{fig:LHC-b}
\end{figure}

\begin{figure}[H]
    \centering
    \includegraphics[width = 1\textwidth]{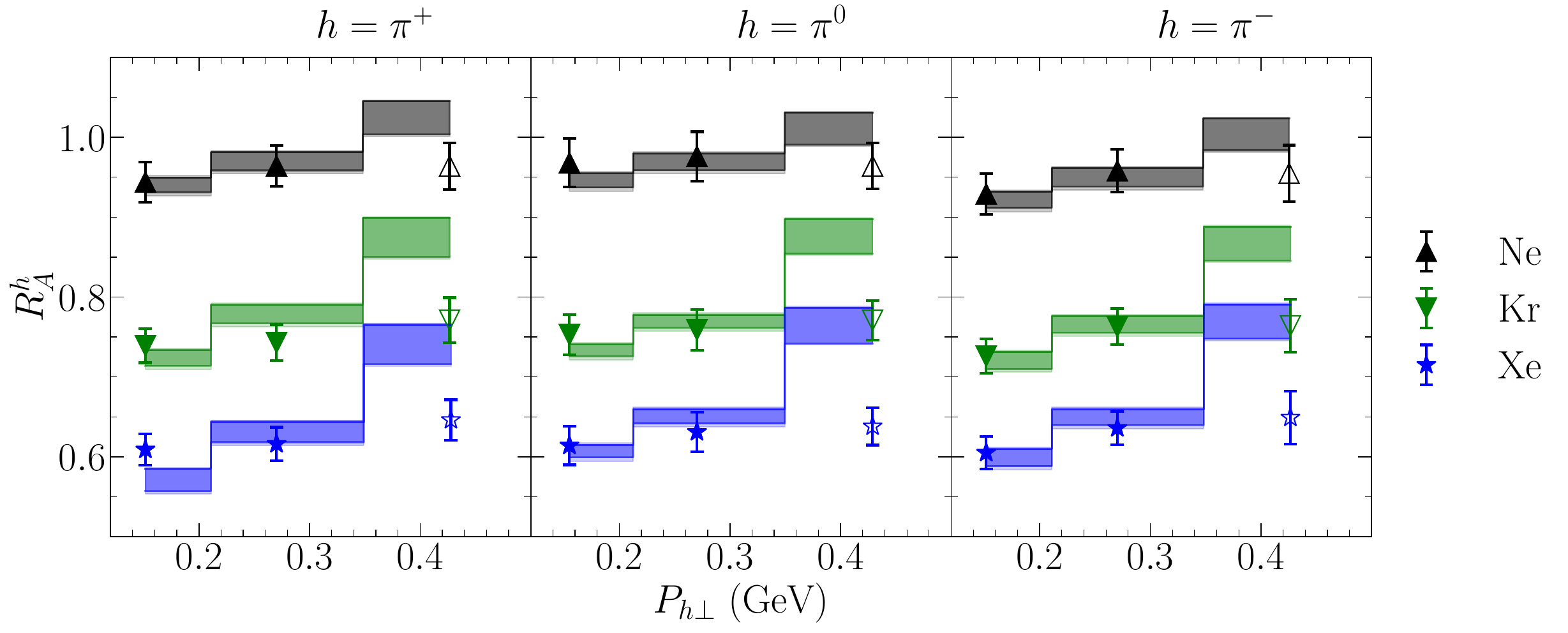}
    \caption{The description of the HERMES dataset. The Xe data has been offset downward by 0.07, while the Ne data has been offset upward by 0.07.}
    \label{fig:HERMES-b}
\end{figure}

\begin{figure}[H]
    \centering
    \includegraphics[width = 1\textwidth]{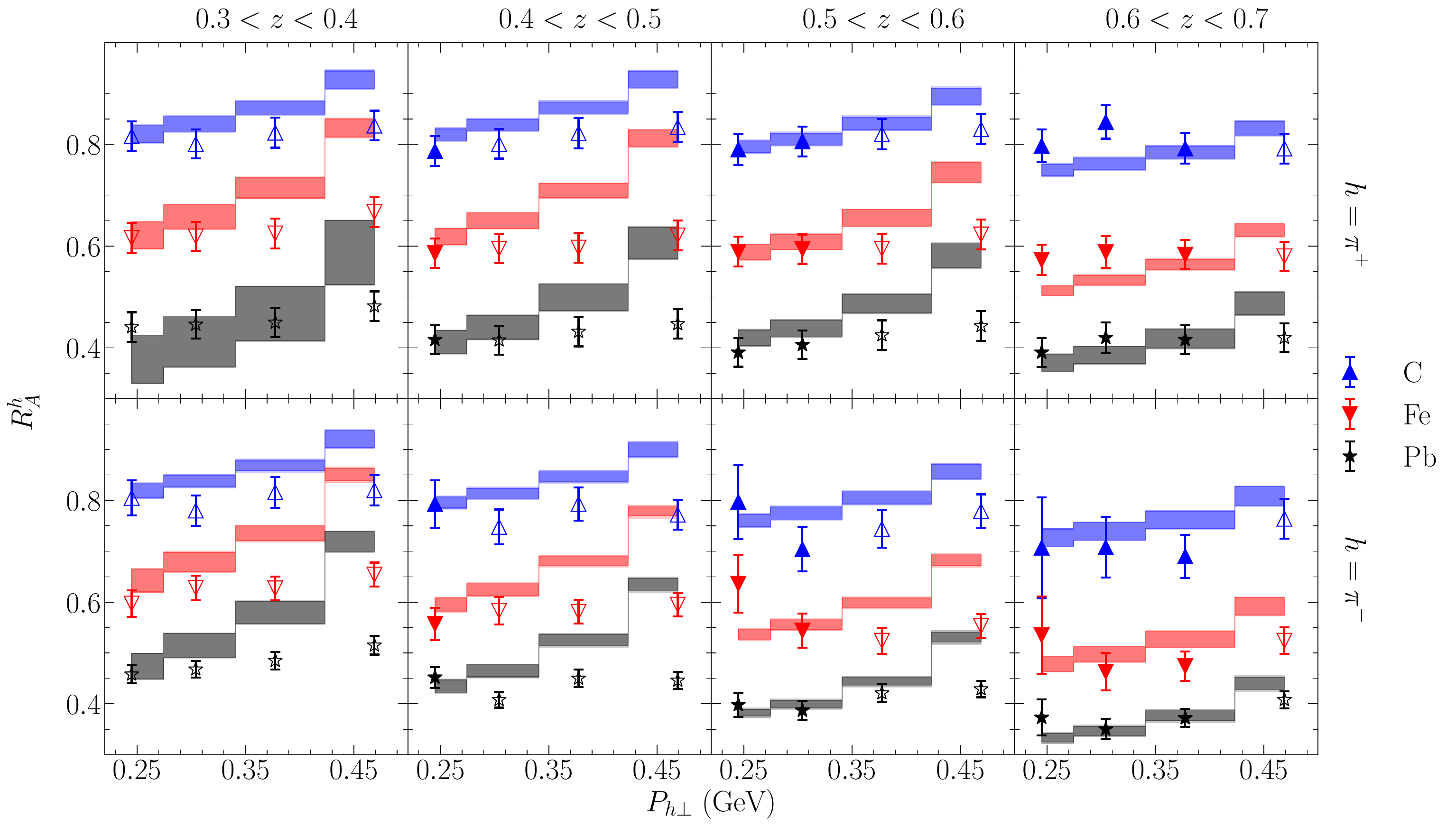}
    \caption{ Theoretical description of JLAB. The dark band represents the fit uncertainty while the light band represents the uncertainty from the PDF.}
    \label{fig:JLab-b}
\end{figure}

In Figs.~\ref{fig:E772-RHIC-b}, \ref{fig:E866-b}, \ref{fig:LHC-b}, \ref{fig:HERMES-b}, and \ref{fig:JLab-b}, we plot the description of the experimental data using the parameter values from fit(b). The data sets in Figs.~\ref{fig:E772-RHIC-b}, \ref{fig:E866-b}, and \ref{fig:LHC-b} are identical to those in fit(a). However in Fig.~\ref{fig:HERMES-b}, we plot the description of the $P_{h\perp}$ projection of the HERMES data. In that plot, the open dots represent experimental data which were not included in the fitting procedure while the solid dots represent those that were. In Fig.~\ref{fig:JLab-b}, we plot the description of the JLab multiplicity ratio. Once again, we use open dots to represent a prediction while solid dots represent data in the fit. In all cases, we find a strong description of the experimental data. 


\begin{figure}[H]
    \centering
    \includegraphics[width = 0.5\textwidth]{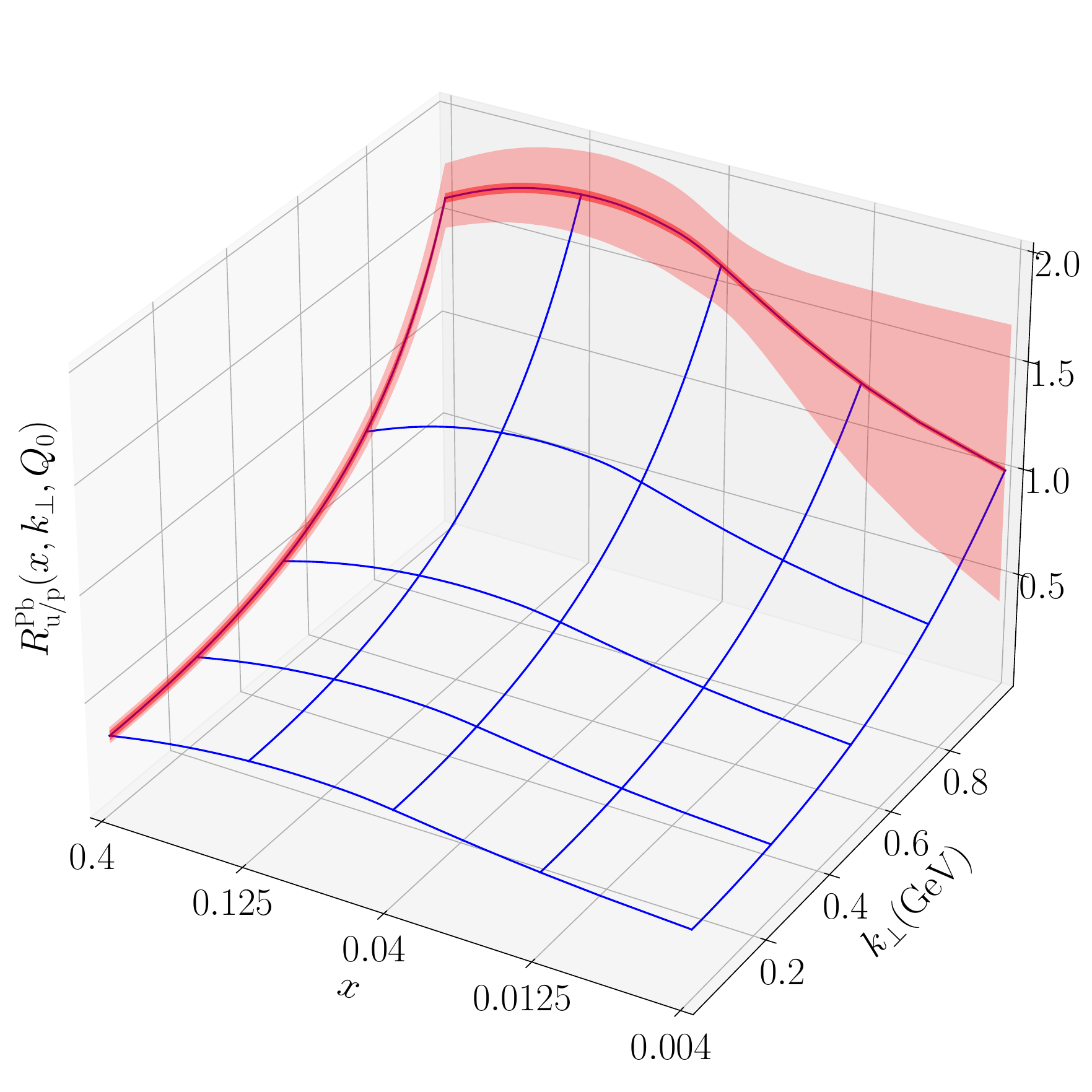} 
    \caption{The extracted nuclear ratio for the TMD FF at $Q_0 = \sqrt{2.4}$ GeV.}
    \label{fig:3D-b}
\end{figure}
\begin{figure}[H]
    \centering
    \includegraphics[width = 1\textwidth]{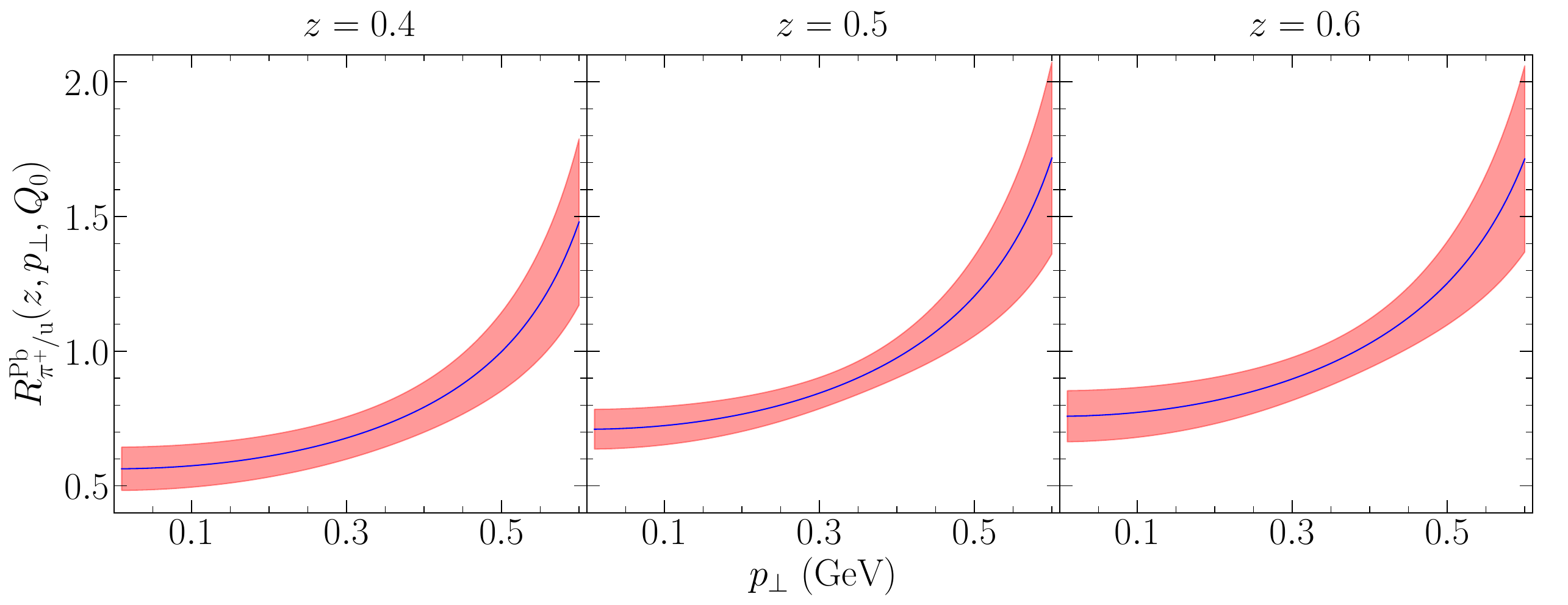} 
    \caption{The extracted nuclear ratio for the TMD FF (down) at $Q_0 = \sqrt{2.4}$ GeV.} 
    \label{fig:nFF-b}
\end{figure}
In Fig.~\ref{fig:3D-b}, once again plot the ratio $R_{u/p}^{\rm Pb}$ at the initial scale. By comparing Figs.~\ref{fig:3Dfita-a} and Fig.~\ref{fig:3D-b}, we see that at the initial scale, the two extractions are consistent with one another. As the range of $z$ values for fit(b) is much more limited than for fit(a) instead of generating a three-dimensional plot, we project the three-dimensional plot onto three curves with different $z$ values in Fig.~\ref{fig:nFF-b}. When studying the $p_\perp$ dependence, we once again see the broadening of the nTMD FF. However, we find for this parameterization that there is no apparent suppression of the nFF at large $z$. This stems from the fact that fit(b) contains data only for the range of $0.4-0.6$. For both the HERMES and JLab data sets, the data is flat in $z$ in that region and thus fit(b) is insensitive to the suppression at larger values of $z$. This issue stems from the limited available data. However future experimental measurements at JLab, the EIC, and the EICC can help to further constrain this behavior.

\section{Predictions}\label{sec:Preds}
\begin{figure}[H]
    \centering
    \includegraphics[width = 1\textwidth]{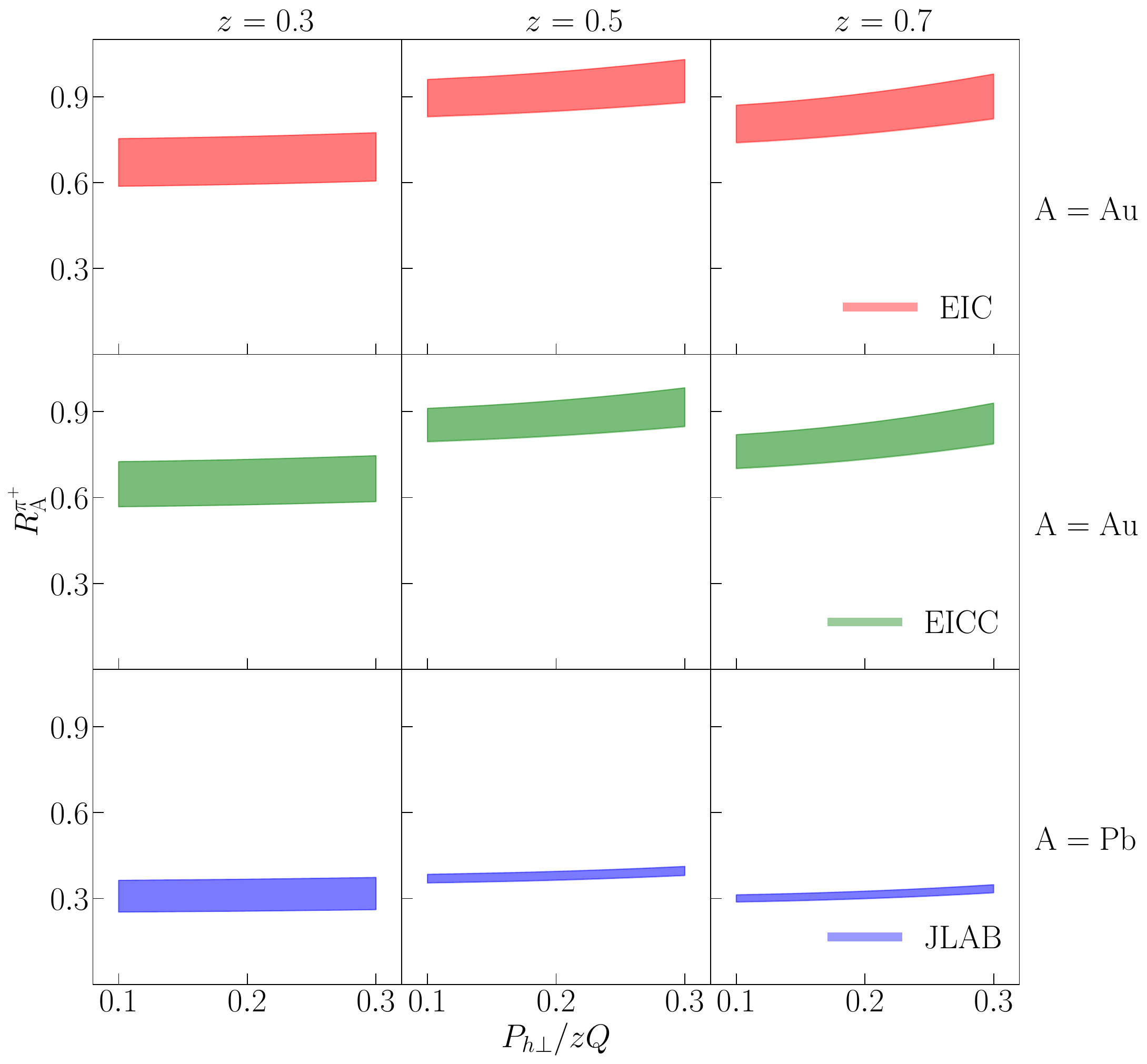}
    \caption{Prediction for multiplicity ratio of EIC, EICC, and JLAB. The band represents the size of the uncertainties.}
    \label{fig:prediction}
\end{figure}
Figure~\ref{fig:prediction} illustrates our projected multiplicity ratios for $\pi^+$ production at various facilities, namely the EIC, EICC, and Jefferson Lab. The predictions are generated using the parameter values from fit (b). In the top row, we present the prediction for the EIC under the configuration $E_\ell = 5$ GeV, $E_N = Z/A\, 41$ GeV, with specific kinematic values $x = 0.1$, $y = 0.2$, corresponding to $Q = 6.58$ GeV$^2$. Moving to the second row, we depict the projection at the EICC with $E_\ell = 3.5$ GeV, $E_N = Z/A\, 20$ GeV, and chosen kinematics $x = 0.1$, $y = 0.5$, resulting in $Q^2 = 5.61$ GeV$^2$. The last row showcases the prediction for $\pi^+$ production in a $\rm{Pb}$ target at Jefferson Lab, employing a $12$ GeV electron beam, and selecting $x = 0.1$, $y = 0.5$. In each plot, the predictions are presented for three values of $z$—$0.3$, $0.5$, and $0.7$—with $P_{h\perp}/zQ$ as the variable.

The plots reveal a distinctive pattern where, at small $P_{h\perp}/zQ$, the cross ratio is suppressed, indicative of a constriction, while larger values signify broadening. It is noteworthy that the prediction at $z = 0.5$ is constrained in our analysis. However, the projections at $z = 0.3$ and $z = 0.7$ fall outside the current range covered by our global analysis. Future experimental data in these regions would be highly valuable for refining and constraining the nFFs in our model. The observed trends underscore the importance of future experimental efforts to precisely determine the behavior of these nFF.

\section{Conclusions}\label{sec:Conclusions}
In this study, we have extended the results of our previous analysis \cite{Alrashed:2021csd} to simultaneously consider nuclear modifications to the nTMDs and nFF. To accomplish this, we have performed the first simultaneous global analysis of nuclear-modified transverse momentum distribution functions (nTMDs) and nuclear-modified collinear fragmentation functions (nFFs) using the global set of experimental data in SIDIS and Drell-Yan. In this paper, we have assessed the impact of the recent Jefferson Lab data by employing two fitting procedures. In the initial fit, a simultaneous analysis of the HERMES and Drell-Yan datasets is performed. In this fit, we find that the extracted nFFs are consistent with the existing LIKEn nFFs, which also rely on the HERMES multiplicity data. In the second fit, we incorporate the Jefferson Lab data and employ the $P_{h\perp}$ projection of the HERMES data. We find that to simultaneously describe the HERMES and Jefferson Lab measurements, we must introduce an additional parameter which characterizes the non-perturbative scale evolution of the nTMD FF. Additionally, we discuss the limitations of the current experimental datasets, offer predictions for future experiments at Jefferson Lab and Electron-Ion Colliders (EICs), and underscore the significance of this work in advancing three-dimensional imaging of nuclear matter and extracting non-perturbative modifications to nFFs. Future work that builds on the progress of this paper may explore the removal of assumptions perturbatively matching the nTMD PDFs onto the nPDFs, extending the methodology of coherent, incoherent multiple scattering and medium induced radiative corrections, such as those in \cite{Qiu:2003vd,Vitev:2003xu,Li:2020zbk,Neufeld:2010dz}, to formulated the TMD dependent DIS and DY cross section, employing a more formal treatment of perturbative interactions in a Glauber SCET framework, and consider experimental measurements from RHIC and the LHC as an avenue into exploring gluon nTMD FFs.

\section*{Acknowledgements}
We thank Daniele Anderle for collaboration at the early stage of the work. The authors thank Christine Aidala, Miguel Arratia, Wei-Yao Ke, and Ivan Vitev for useful discussions. We also thank Hannu Paukkunen and the other authors of the EPPS21 analysis for providing the Kr and Xe grids. H.X. is supported by the Guangdong Major Project of Basic and Applied Basic Research No. 2020B0301030008, the Key Project of Science and Technology of Guangzhou (Grant No. 2019050001), the National Natural Science Foundation of China under Grant No. 12022512, No.~12035007. M.A. is supported by the Kuwait University Graduate Scholarship. Z.K. is supported by the National Science Foundation under Grant No.~PHY-1945471. J.T. is supported by the Department of Energy at LANL through the LANL/LDRD Program under project number~20220715PRD1. C.Z. is supported by the UCLA Physics REU program. This work is also supported by the U.S. Department of Energy, Office of Science, Office of Nuclear Physics, within the framework of the Saturated Glue (SURGE) Topical Theory Collaboration.

\appendix

\section{One loop expressions}
We write the full expressions for the Wilson coefficient functions as a perturbative series
\begin{align}
    C_{i\leftarrow j}(x,\mu_b,\mu,\zeta) = \sum_i \left(\frac{\alpha_s}{4\pi}\right)^i C_{f\leftarrow f'}^{(i)}(x,\mu_b,\mu,\zeta)\,,
\end{align}
where we have written this explicitly for the unpolarized TMD PDF but we note that we take this labeling convention for all TMDs. 

At LO, the only non-zero matching coefficients are given by
\begin{align}
    C_{q\leftarrow q}^{(0)}(x,\mu_b,\mu,\zeta) & = \delta\left(1-x\right)
    \\
    C_{g\leftarrow g}^{(0)}(x,\mu_b,\mu,\zeta) & = \delta\left(1-x\right)\,,
\end{align}
where we have included the gluon matching for completeness. At one loop order, the matching functions are given for instance in \cite{Echevarria:2016scs} as
\begin{align}
    C_{q\leftarrow q}^{(1)}(x,\mu_b,\mu,\zeta) & = C_F\left[-2 L_\mu p_{qq}(x)+2(1-x)+\delta\left(1-x\right)\left(-L_\mu^2+2L_\mu L_\zeta-\frac{\pi^2}{6}\right)\right]\,, \nn \\
    C_{q\leftarrow g}^{(1)}(x,\mu_b,\mu,\zeta) & = T_r\left[-2L_\mu p_{gq}(x)+4 x(1-x)\right]\,, \nn \\
    \label{eq:match}
    C_{g\leftarrow q}^{(1)}(x,\mu_b,\mu,\zeta) & = C_F\left[-2L_\mu p_{qg}(x)+2x\right]\,, \nn \\
    C_{g\leftarrow g}^{(1)}(x,\mu_b,\mu,\zeta) & = C_A\bqty{-4 L_\mu p_{gg}(x)+\delta\pqty{1-x} \pqty{-L_\mu^2+2 L_\mu L_\zeta -\frac{\pi^2}{6}}}\,,
\end{align}
where the logarithms are defined as
\begin{align}
    L_\mu = \log\left(\frac{\mu^2}{\mu_b^2}\right)\,,
    \qquad
    L_\zeta = \log\left(\frac{\mu^2}{\zeta}\right)\,.
\end{align}
In this expression, we have introduced the collinear splitting functions which are given by
\begin{align}
    P_{qq}(x) &= \frac{1+x^2}{(1-x)_+}+\frac{3}{2}\delta\left(1-x\right)\,, \nn \\
    P_{qg}(x) &= 1-2x(1-x)\,, \nn \\
    P_{gq}(x) &= \frac{1+(1-x)^2}{x}\,, \nn \\
    \qquad
    P_{gg}(x) &= \frac{(1-x(1-x))^2}{x(1-x)_+}+\delta\pqty{1-x} \frac{11 C_A-4 T_R}{6 C_A}\,.
\end{align}
\begin{align}
    p_{qq}(x) &= \frac{1+x^2}{(1-x)_+}\,, \nn \\
    p_{qg}(x) &= 1-2x(1-x)\,, \nn \\
    p_{gq}(x) &= \frac{1+(1-x)^2}{x}\,, \nn \\
    \qquad
    p_{gg}(x) &= \frac{(1-x(1-x))^2}{x(1-x)_+}\,.
\end{align}
The matching coefficients for the unpolarized TMD FFs can be related to those of the collinear TMD PDFs through the relations
\begin{align}
    \hat{C}_{q\leftarrow q}(z,\mu_b,\mu,\zeta) &= C_{q\leftarrow q}(z,z\,\mu_b,\mu,\zeta)  \nn \\
    \hat{C}_{q\leftarrow g}(z,\mu_b,\mu,\zeta) &= C_{g\leftarrow q}(z,z\,\mu_b,\mu,\zeta) \nn \\
    \hat{C}_{g\leftarrow q}(z,\mu_b,\mu,\zeta) &= C_{q\leftarrow g}(z,z\,\mu_b,\mu,\zeta)\nn \\
    \hat{C}_{g\leftarrow g}(z,\mu_b,\mu,\zeta) &= C_{g\leftarrow g}(z,z\,\mu_b,\mu,\zeta)\,,
\end{align}
which holds at least up to NLO. We emphasize that the natural scale $\mu_b$ of the TMD FFs differs from those of the TMD PDFs, and thus gives rise to the threshold logarithms in the expressions for the TMD FFs such that $L_\mu \rightarrow L_\mu-\ln{z}^2$ in the expressions for the TMD FFs.

Lastly, we provide the expressions for the one-loop hard function for DIS and Drell-Yan
\begin{align}\label{eq:hard-DIS}
    H_{\rm DIS}(Q,\mu) & = 1+\frac{\alpha_sC_F}{2\pi}\left[3 L_Q-L_Q^2-8+\frac{\pi^2}{6}\right]\,,
    \\
    \label{eq:hard-DY}
    H_{\rm DY}(Q,\mu) & = 1+\frac{\alpha_sC_F}{2\pi}\left[3 L_Q-L_Q^2-8+\frac{7\pi^2}{6}\right]\,,
\end{align}
where the logarithms are given by $L_Q = \log\left(Q^2/\mu^2\right)$.
\section{Anomalous dimensions up to NNLL}
The anomalous dimensions of the hard function, TMD PDF, and TMD FF are given by
\begin{align}\label{eq:anom-hard}
    \gamma^H_\mu(\mu) & = 2C_F\gamma^{\rm cusp}\left[\alpha_s(\mu)\right] \ln\left(\frac{Q^2}{\mu^2}\right)+4\gamma_q\left[\alpha_s(\mu)\right]
    \\
    \label{eq:anom-f}
    \gamma^f_\mu(\mu,\zeta) & = -C_F\gamma^{\rm cusp}\left[\alpha_s(\mu)\right]\ln\left(\frac{\zeta}{\mu^2}\right)-2\gamma_q\left[\alpha_s(\mu)\right]
    \\
    \label{eq:anom-D}
    \gamma^D_\mu(\mu,\zeta) & = -C_F\gamma^{\rm cusp}\left[\alpha_s(\mu)\right]\ln\left(\frac{\zeta}{\mu^2}\right)-2\gamma_q\left[\alpha_s(\mu)\right]\,.
\end{align}
In this expression, $\gamma_{\rm cusp}$ and $\gamma_q$ are the non-cusp anomalous dimension. These can be expressed as a perturbative in the strong coupling as
\begin{align}
    \gamma^{\rm cusp}\left[\alpha_s(\mu)\right] = \sum_{i = 0}^\infty \left(\frac{\alpha_s}{4\pi}\right)^{i+1}\gamma^{\rm cusp}_i
    \qquad
    \gamma^q\left[\alpha_s(\mu)\right] = \sum_{i = 0}^\infty \left(\frac{\alpha_s}{4\pi}\right)^{i+1}\gamma^q_i\,.
\end{align}
At NNLL, the cusp and non-cusp terms are given by~\cite{Korchemsky:1987wg, Moch:2004pa,Moch:2005id,Moch:2005tm,Idilbi:2005ni,Idilbi:2006dg,Becher:2006mr}
\begin{align}
\label{eq:c1}
  \gamma_{0}^{\rm cusp} = & \, 4 \,,
  \\
  \gamma_{1}^{\rm cusp} = & \, C_A  \left(\frac{268}{9}-8 \zeta_2\right)-\frac{40  n_f}{9} \nn \,,
  \\
  \gamma_{2}^{\rm cusp} = & \, C_A^2  \left(-\frac{1072 \zeta_2}{9}+\frac{88
      \zeta_3}{3}+88 \zeta_4+\frac{490}{3}\right) \nn \\
    & +C_A  n_f \left(\frac{160 \zeta_2}{9}-\frac{112
    \zeta_3}{3}-\frac{836}{27}\right) \nn \\
    & + C_F n_f \left(32 \zeta_3-\frac{110}{3}\right)-\frac{16
                 n_f^2}{27} \,.
\nn
\end{align}
\begin{align}
  \gamma^{q}_0 = & \, -3 C_F \,, \\
  \gamma^{q}_1 = & \, C_A C_F \left(-11 \zeta_2+26
    \zeta_3-\frac{961}{54}\right) \nn \\
& +C_F^2 \left(12 \zeta_2-24 \zeta_3-\frac{3}{2}\right)+C_F n_f \left(2 \zeta_2+\frac{65}{27}\right) \,, \nn \\
\end{align}
Similarly, the Collins-Soper anomalous dimension of the TMDs can be written as
\begin{align}\label{eq:anom-zeta}
    \gamma_\zeta(\mu,b) = -2 C_F \int_{\mu_b}^\mu \frac{d\mu'}{\mu'}\Gamma_{\rm cusp}\left[\alpha_s\left(\mu'\right)\right]-C_F \gamma^r\left[\alpha_s(\mu_b)\right]\,,
\end{align}
where $\gamma^r$ is the rapidity anomalous dimension that is known up to four loops~\cite{Duhr:2022yyp,Moult:2022xzt}. This anomalous dimension can be expressed at a perturbative series as
\begin{align}
    \gamma^r\left[\alpha_s(\mu)\right] = \sum_{i = 0}^\infty \left(\frac{\alpha_s}{4\pi}\right)^{i+1}\gamma^r_i\,.
\end{align}
where at NNLL, the anomalous dimensions are \cite{Almelid:2015jia,Almelid:2017qju}
\begin{align}
    \gamma_{0}^r = & \, 0 \,, \\
    \gamma_{1}^r = & \, C_A \left(\frac{22 \zeta_2}{3}+28 \zeta_3-\frac{808}{27}\right)+ n_f \left(\frac{112}{27}-\frac{4 \zeta_2}{3}\right)- 2 \zeta_2 \beta_0 \,. \nn
\end{align}

\bibliography{main}

\bibliographystyle{JHEP}
\end{document}